\documentclass[showpacs,preprint,preprintnumbers,showpacs,showkeys,superscriptaddress,amsmath,amssymb,nofootinbib]{revtex4}
\usepackage{amsmath}
\usepackage{latexsym}
\usepackage{pstricks,pst-coil}
\usepackage{pst-all}

\renewcommand{\theequation}{\thesection.\arabic{equation}}
\usepackage{color}
\usepackage{latexsym}
\usepackage{amsmath}
\usepackage{amssymb}
\usepackage{eufrak}
\usepackage{euscript}
\usepackage{pstricks}
\usepackage{graphics}
\usepackage{graphicx}
\usepackage{picture}

\newcommand{\be}{\begin{equation}}
\newcommand{\ee}{\end{equation}}
\newcommand{\ba}{\begin{eqnarray}}
\newcommand{\ea}{\end{eqnarray}}
\newcommand{\p}{\partial}
\def\ni{\noindent}

\def\uma{\rm 1\!\!\hskip 1 pt l}

\begin{document}

\title{\Large{The standard electroweak model in the noncommutative $DFR$ space-time}}

\author{M. J. Neves}\email{mariojr@ufrrj.br}
\affiliation{Grupo de F\'{i}sica Te\'orica e Matem\'atica F\'{i}sica, Departamento de F\'{\i}sica, Universidade Federal Rural do Rio de Janeiro, BR 465-07, 23890-971, Serop\'edica, RJ, Brazil}
\author{Everton M. C. Abreu}\email{evertonabreu@ufrrj.br}
\affiliation{Grupo de F\'{i}sica Te\'orica e Matem\'atica F\'{i}sica, Departamento de F\'{\i}sica, Universidade Federal Rural do Rio de Janeiro, BR 465-07, 23890-971, Serop\'edica, RJ, Brazil}
\affiliation{Departamento de F\'{i}sica, Universidade Federal de Juiz de Fora, 36036-330, Juiz de Fora, MG, Brazil}

\date{\today}


\begin{abstract}
\noindent The noncommutative (NC) framework elaborated by Doplicher, Fredenhagen and Roberts (DFR) has a Lorentz invariant spacetime structure in order to be considered as a candidate to understand the physics of the early Universe.  In DFR formalism the NC parameter ($\theta^{\mu\nu}$) is a coordinate operator in an extended Hilbert space and it has a conjugate momentum.  Since $x$ and $\theta^{\mu\nu}$ are independent coordinates, the Weyl-Moyal (WM) product can be used in this framework.  With these elements, in this work, we have constructed the standard electroweak model.  To accomplish this task we have begun with the construction of a DFR metric tensor.  After that we have analyzed the
WM-product basis group of symmetry.  After that we have introduced the spontaneous symmetry breaking and the hypercharge in DFR framework.  The electroweak  symmetry breaking was analyzed and the masses of the new bosons were computed.  Finally, the gauge symmetry and gauge transformations were discussed.
\end{abstract}



\maketitle

\pagestyle{myheadings}
\markright{The standard electroweak model in the non-commutative $DFR$ space-time}


\section{Introduction}
\renewcommand{\theequation}{1.\arabic{equation}}
\setcounter{equation}{0}

The fact that we have infinities that destroy the final results of several calculations in QFT has motivated theoretical physicists to ask if a continuum space-time would be really necessary. One of the possible solutions would be to create a discrete space-time with a noncommutative (NC) algebra,
where the position coordinates would be promoted to operators $\hat{X}^{\mu}\,(\mu=0,1,2,3)$ and they must satisfy commutation relations
\be\label{xmuxnu}
\left[\,\hat{X}^\mu\,,\,\hat{X}^\nu\,\right]\,=\,i\,\ell \, \theta^{\mu\nu} \, \hat{\openone} \,\,,
\ee
where $\ell$ is a length parameter, $\theta^{\mu\nu}$ is an antisymmetric constant matrix and $\hat{\openone}$ is the identity operator.  In this way, we would have a kind of fuzzy space-time where, from (\ref{xmuxnu}) we have an uncertainty in the position coordinate.
In order to put these ideas together, Snyder \cite{snyder47} have published the first work that considers the space-time as being NC. However, C. N. Yang \cite{yang47} little time latter have demonstrated that Snyder's dream about freeing QFT from the infinities were not achieved by noncommutativity (NCY).  This result have doomed Snyder's NC theory to years of ostracism.   After the relevant result that the algebra obtained from string theory embedded in a magnetic background is NC, a new flame concerning NCY was rekindle \cite{seibergwitten99}.

One of the paths (the most famous at least) of introducing NCY is through the Moyal-Weyl product where the NC parameter,
{\it i.e.}, $\theta^{\mu\nu}$, is an antisymmetric constant matrix. However, at superior orders of calculations, the Moyal-Weyl product turns out to be
highly nonlocal. This fact forced us to work with low orders in $\theta^{\mu\nu}$. Although it keeps the translational invariance,
the Lorentz symmetry is not preserved \cite{Szabo03}. For instance, concerning the case of the hydrogen atom, it breaks the rotational symmetry of the model, which removes the degeneracy of the energy levels \cite{Chaichian}.  Other subjects where the objective was to introduce NC effects in gravity \cite{grav}, in anyon models \cite{anyons} and symmetries \cite{sym}.  NCY through path integrals and coherent states was devised in \cite{path}.  For more generalized NC issues and reviews, the interested reader can look at \cite{reviews} and the references therein.

To fix this Lorentz symmetry problem, in order to work with the NCY space-time was introduced by Doplicher, Fredenhagen and Roberts.   They have considered the parameter $\theta^{\mu\nu}$ as an ordinary coordinate of the system \cite{DFR1,DFR2}. This extended and new NC space-time has originally ten dimensions: four relative to the Minkowski space-time and six relative to $\theta$-space. Recently, in \cite{Morita,alexei,jhep,Amorim1,saxell,Amorim4,Amorim5,Amorim2} the authors have demonstrated that the DFR formalism have in fact a canonical momentum associated with $\theta^{\mu\nu}$
\cite{alexei,saxell,Amorim1} (for a review the interested reader can see \cite{amo}). The DFR framework is characterized by a field theory constructed in a space-time with extra-dimensions $(4+6)$,
and which does not need necessarily the presence of a length scale $\ell$ localized into the six dimensions
of the $\theta$-space, where, from (\ref{xmuxnu}) we can see that $\theta^{\mu\nu}$ has dimension of length-square,
when we make $\ell=1$. By taking the limit with no such scale, the usual algebra of
the commutative space-time is recovered. Besides the Lorentz invariance was recovered, we obviously hope that causality aspects in QFT in this $\left(x+\theta\right)$ space-time must be preserved too.

In the last few years, we have been concentrated in the construction of a NC QFT concerning the DFR formalism \cite{EMCAbreuMJNeves2012,EMCAbreuMJNeves2013,EMCAbreuMJNeves2014,NevesAbreuEPL}.  In this work we, following this DFR QFT quest, have constructed the DFR basis for a standard model after the construction of a DFR metric tensor.

The organization of this paper follows the schedule: in section 2 we have reviewed the DFR  NC formalism.  The main commutator relations were depicted.
In section 3 we have constructed a metric tensor where the unitarity of the system was an issue.
In section 4, we have constructed the DFR NC version of the Glashow-Salam-Weinberg (GSW) model for the electroweak interaction.
In section 5 we have introduced the first Higgs sector to identify the hypercharge and we have analyzed the spontaneous symmetry breaking.   The 3-line and 4-line vertex interactions were discussed.
In section 6 we have discussed the electroweak symmetry breaking combined to the definition of a second Higgs sector.   In  section 7, we have brought some DFR numerical results relative to the masses of the new bosons.   Interactions between leptons and neutrino fields were studied and NC interactions Lagrangians were constructed.
Finally, the gauge symmetry were discussed in section 8 and the conclusions were depicted in section 9.


\section{The DFR NC space-time and Field Theory}
\renewcommand{\theequation}{2.\arabic{equation}}
\setcounter{equation}{0}

In this section, we will construct a NC model based on the results published in \cite{Amorim1,saxell,Amorim4,Amorim4,Amorim5,Amorim2,amo}.
Namely, we will revisit the basics of QFT analogous to the one defined in DFR space.
As we have said before, in DFR formalism the parameters $\theta^{\mu\nu}$ are promoted
to coordinate-operator in this space-time, which has $D=10$, it has six independents spatial coordinates,
which are, $\theta^{\mu\nu}=\left(\theta^{01},\theta^{02},\theta^{03},\theta^{12},\theta^{13},\theta^{23}\right)$.
To avoid unitarity problems, we will use that the time components of $\theta^{\mu\nu}$ are zero, namely, $\theta^{0i}=0$, and the NC approach is
purely spatial $\theta^{\mu\nu}=(0,\theta^{ij})$.   Consequently, we have a space-time of $D=7$.  In the next section we will construct a metric tensor for the DFR space.
The spatial coordinates
of $\theta^{\mu\nu}$ are promoted to quantum observable $\hat{\Theta}^{ij}$ in the commutation relation (\ref{xmuxnu}),
so the spatial non-commutativity is given by
\begin{eqnarray}\label{algebraDFR}
\left[\hat{X}^{i},\hat{X}^{j}\right] = i \, \hat{\Theta}^{ij} \; ,
\end{eqnarray}
while the time operator does commute with spatial operator usually
\begin{eqnarray}
\left[\hat{X}^{0},\hat{X}^{i}\right] = 0 \; .
\end{eqnarray}
By convenience, we use the dual operator of $\hat{\Theta}^{ij}$ as our NC
coordinate $\hat{\Theta}^{ij}=\epsilon^{ \, ijk} \, \hat{\Theta}^{k}$.
Moreover there exist the canonical conjugate momenta operator $\hat{K}^{ij}=\epsilon^{\, ijk} \, \hat{K}^{k}$ associated \footnote{The standard notation to represent the momentum is $\pi_{\mu\nu}$ but, for future to-avoid-confusion convenience, we will use from now on, $\hat{K}^{\mu\nu}$.} with the operator $\hat{\Theta}^{ij}$, and they must satisfy the commutation relation
\begin{equation}\label{thetapicomm}
\left[\,\hat{\Theta}^{i},\hat{K}^{j}\, \right] = i \, \delta^{ij} \, \hat{\uma} \; .
\end{equation}
%
In order to obtain consistency we can write that \cite{Amorim1}
\begin{eqnarray}\label{algebraDFRextended}
\left[\hat{X}^{\mu},\hat{P}^{\nu} \right] = i\,\eta^{\mu\nu} \, \hat{\uma}
\hspace{0.2cm} , \hspace{0.2cm}
\left[\hat{P}^{\mu},\hat{P}^{\nu} \right] = 0
\hspace{0.2cm} , \hspace{0.2cm}
\left[\hat{\Theta}^{ij},\hat{P}^{\rho}\right] = 0
\hspace{0.2cm} , \hspace{0.2cm}
\nonumber \\
\left[\hat{P}^{\mu},\hat{K}^{ij}\right] = 0
\hspace{0.2cm} , \hspace{0.2cm}
\left[\hat{X}^{i},\hat{K}^{jk}\right]=- \, \frac{i}{4} \, \delta^{ij} \hat{P}^{k} + \frac{i}{4} \, \delta^{ik} \hat{P}^{j}  \; ,
\end{eqnarray}
and these relations complete the DFR extended algebra\footnote{Here we have adopted that $c=\hbar=\ell=1$, where the $\theta$-coordinate has area dimension.}. It is possible to verify that (2.1)-(2.3) commutation relations listed above are indeed consistent with all possible
Jacobi identities and the CCR algebras \cite{EMCAbreuMJNeves2012}.

The uncertainty principle from (\ref{xmuxnu}) is modified by
\begin{eqnarray}\label{uncertainxmu}
\Delta \hat{X}^{i} \Delta \hat{X}^{j} \simeq \langle \hat{\Theta}^{ij}\rangle \; ,
\end{eqnarray}
where the expected value of the operator $\hat{\Theta}$ is related to the fluctuation position
of the particles, and it has dimension of length-squared.

The last commutation relation in Eq. (\ref{algebraDFRextended}) suggests that the shifted coordinate
operator \cite{Chaichian,Gamboa,Kokado,Kijanka,Calmet1,Calmet2}
\begin{equation}\label{X}
\hat{\xi}^{i} = \hat{X}^{i}\,+\,{\frac 12}\,\hat{\Theta}^{ij}\hat{P}^{j} \,\,,
\end{equation}
commutes with $\hat{K}^{ij}$.  The relation (\ref{X}) is also known in the algebraic literature as Bopp shift.
The commutation relation (\ref{algebraDFR}) also commutes with $\hat{\Theta}^{ij}$ and $\hat{\xi}^{i}$,
and satisfies a non trivial commutation relation with $\hat{P}^{\,\mu}$ dependent objects, which could be derived from
\begin{equation}\label{Xpcomm}
[\hat{\xi}^{\mu},\hat{P}^{\nu}]=i\,\eta^{\mu\nu} \, \hat{\uma}
\hspace{0.6cm} , \hspace{0.6cm}
[\hat{\xi}^{\mu},\hat{\xi}^{\nu}]=0\,\, ,
\end{equation}
and we can note that the property $\hat{P}_{\mu} \, \hat{\xi}^{\mu}=\hat{P}_{\mu} \, \hat{X}^{\mu}$ is easily verified.
Hence, we can see from these both equations that the shifted coordinated operator (\ref{X}) allows us to recover
the commutativity. The shifted coordinate operator $\hat{\xi}^{i}$ plays a fundamental role in NC
quantum mechanics defined in the $\left(x+\theta\right)$-space \cite{Amorim1}, since it is possible to form a basis with its eigenvalues.
So, differently from $\hat{X}^{i}$, we can say that $\hat{\xi}^{i}$ forms a basis in
Hilbert space. The framework showed above demonstrated that in NCQM, the physical coordinates
do not commute and the respective eigenvectors cannot be used to form a basis
in ${\cal H}={\cal H}_1 \oplus {\cal H}_2$ \cite{Amorim4}.  This can be accomplished
with the Bopp shift defined in (\ref{X}) with (\ref{Xpcomm}) as consequence.
%
%
%
%
%
%

%
The Lorentz generator group is
\begin{eqnarray}\label{Mmunu}
\hat{M}_{\mu\nu}=\,\hat{\xi}_{\mu}\hat{P}_{\nu}\,-\,\hat{\xi}_{\nu}\hat{P}_{\mu}
+\hat{\Theta}_{\nu\rho}\hat{K}^{\rho}_{\; \;\mu}-\hat{\Theta}_{\mu\rho}\hat{K}^{\rho}_{\; \;\nu} \; ,
\end{eqnarray}
and from (\ref{algebraDFRextended}) we can write the generators
for translations as $\hat{P}_{\mu} \rightarrow i \partial_{\mu}\,\,$.
With these ingredients it is direct to construct the commutation relations
\begin{eqnarray}\label{algebraMP}
\left[ \hat{P}_\mu , \hat{P}_\nu \right] &=& 0
\hspace{0.2cm} , \nonumber \\
\left[ \hat{M}_{\mu\nu},\hat{P}_{\rho} \right] &=& \,i\,\big(\eta_{\mu\rho}\,\hat{P}_\nu
-\eta_{\nu\rho}\,\hat{P}_\mu\big) \; ,
\hspace{0.1cm} \nonumber \\
\left[\hat{M}_{\mu\nu} ,\hat{M}_{\rho\sigma} \right] &=& i\left(\eta_{\mu\sigma}\hat{M}_{\rho\nu}
-\eta_{\nu\sigma}\hat{M}_{\rho\mu}-\eta_{\mu\rho}\hat{M}_{\sigma\nu}+\eta_{\nu\rho}\hat{M}_{\sigma\mu}\right) \; ,
\end{eqnarray}
which closes the proper algebra.  We can say that $\hat{P}_\mu$ and $\hat{M}_{\mu\nu}$
are the DFR algebra generators.

\section{DFR metric tensor and relativistic quantum mechanics}

Considering what we have said in the last section, geometrically speaking we have the structure of a plane space-time $4D$ NC attached to a $3D$ extra-dimensional spatial coordinates $\{ \theta^{i} \}$
compactified in a sphere $S^{2}$ which square ratio is conjectured by the length scale of the non-commutativity. We construct the measure of this space-time by means of the line element
\begin{eqnarray}
ds^{2}=\eta_{\mu\nu} \, dx^{\mu} \, dx^{\nu}+\frac{e^{-\frac{\theta_{1}^{\,2}}{4\lambda^{4}}}}{\lambda^{2}} \, \left(d\theta_{1} \right)^{2}
+\frac{e^{-\frac{\theta_{2}^{\,2}}{4\lambda^{4}}}}{\lambda^{2}} \, \left(d\theta_{2} \right)^{2}
+\frac{e^{-\frac{\theta_{3}^{\,2}}{4\lambda^{4}}}}{\lambda^{2}} \, \left(d\theta_{3} \right)^{2} \; ,
\end{eqnarray}
where $\eta^{\mu\nu}=\mbox{diag}(1,-1,-1,-1)$ is the usual Minkowski metric. The sector of extra-dimension is compactified by
gaussian functions that keep the isotropy of the $\theta$-space. Here we have a matrix $7 \times 7$ as the metric of a $4+3$
space-time
\begin{eqnarray}\label{metricg}
G\,=\,\mbox{diag}\left(1 \, , \, -1 \, , \, -1 \, , \, -1 \, , \, e^{-\frac{\theta_{1}^{\, 2}}{4\lambda^{4}}} \, , \, e^{-\frac{\theta_{2}^{\, 2}}{4\lambda^{4}}}
\, , \, e^{-\frac{\theta_{3}^{\, 2}}{4\lambda^{4}}} \right) \; .
\end{eqnarray}
An important point in DFR algebra issue is that the Weyl representation of NC operators obeying
the commutation relations keeps the usual form of the Moyal product. In this case, the Weyl map
is represented by
\begin{eqnarray}\label{mapweyl}
\hat{{\cal W}}(f)(\hat{X},\hat{\Theta})=\int\frac{d^{4}p}{(2\pi)^{4}}
\frac{d^{3}{\bf k}}{(2\pi\lambda^{-2})^{3}} \;
\widetilde{f}(p,k) \; e^{i \, p_{\mu} \hat{X}^{\mu} \, + \, i \, {\bf k} \, \cdot \, \hat{\Theta}} \; .
\end{eqnarray}
The Weyl symbol provides a map from the operator algebra to the functions algebra equipped with a star-product,
via the Weyl-Moyal correspondence
\begin{eqnarray}\label{WeylMoyal}
\hat{f}(\hat{X},\hat{\Theta}) \; \hat{g}(\hat{X},\hat{\Theta})
\hspace{0.3cm} \longleftrightarrow \hspace{0.3cm}
f(x,\theta) \star g(x,\theta) \; ,
\end{eqnarray}
where the star-product $\star$ is defined by
\begin{eqnarray}\label{ProductMoyal}
\left. f(x,\theta) \star g(x,\theta) =
e^{\frac{i}{2}\theta^{ij}\partial_{i}\partial^{\prime}_{j}}
f(x,\theta) \, \, g(x^{\prime},\theta) \right|_{x^{\prime}=x} \; ,
\end{eqnarray}
for any arbitrary functions $f$ and $g$ of the coordinates $(x^{\mu},\theta^{ij})$.
Namely, in both sides of Eq. (\ref{ProductMoyal}) we have that $f$ and $g$ are NC objects since they depend on $\theta^{ij}$.

The Weyl operator (\ref{mapweyl}) has the trace property considering a product of $n$ NC functions $(f_{1},...,f_{n})$
\begin{eqnarray}\label{traceWfs}
\mbox{Tr}\left[ \hat{{\cal W}}(f_{1}) ... \hat{{\cal W}}(f_{n}) \right]=
\int d^{4}x \int \frac{d^{3}\theta}{(2\pi \lambda^{2})^{3}} \; \sqrt{-g} \; f_{1}(x,\theta) \star ... \star f_{n}(x,\theta) \; ,
\end{eqnarray}
where $g$ inside the integral means the determinant of the metric (\ref{metricg}). The calculus of this determinant only depends on
the $\theta^{i}$-coordinates, so the integral can be written as
\begin{eqnarray}\label{traceWfs}
\mbox{Tr}\left[ \hat{{\cal W}}(f_{1}) ... \hat{{\cal W}}(f_{n}) \right]=
\int d^{4}x \; d^{3}\theta \; W(\theta) \; f_{1}(x,\theta) \star ... \star f_{n}(x,\theta) \; ,
\end{eqnarray}
where the $W$ normalized function is defined by
\begin{eqnarray}
W(\theta_{i})=\left(\frac{1}{2\pi\lambda^{2}} \right)^{3} \, e^{-\frac{\theta_{i} \, \theta_{i}}{4\lambda^{4}}} \; .
\end{eqnarray}

The function $W$ is a $\theta$-integration measure that we have introduced thanks to the geometry of the $4+7$ dimensional space-time.
This weight function is introduced in the context of NC field theory to control divergences of the integration
in the $\theta$-space \cite{Carlson,Morita,Conroy2003,Saxell}. Theoretically speaking, it would permit us to work with series expansions in $\theta$, {\it i.e.}, with truncated power series expansion of functions of $\theta$.
For any large $\theta^{i}$ it falls to zero quickly so that all integrals are well defined,
in its definitions the normalization condition was assumed when integrated in the $\theta$-space.
The function $W$ should be an even function of $\theta$, that is, $W(-\theta)=W(\theta)$ which implies that
an integration in $\theta$-space be isotropic. All the properties involving the
$W$-function can be seen in details in \cite{Carlson,Morita,Conroy2003,Saxell}.
However, we have to say that the role of the $W$-function in NC issues is not altogether clear among the NC community.
By the definition of the Moyal product (\ref{ProductMoyal}) it is
trivial to obtain the property
\begin{eqnarray}\label{Idmoyalproduct2}
\int d^{4}x\,d^{3}\theta \; W(\theta) \; f(x,\theta) \star g(x,\theta)=
\int d^{4}x\,d^{3}\theta \; W(\theta) \; f(x,\theta) \; g(x,\theta) \; .
\end{eqnarray}
%
%
%
%
The physical interpretation of the average of the components of $\theta^{ij}$, {\it i.e.}
$\langle \theta^{2} \rangle$, is the definition of the NC energy scale \cite{Carlson}
\begin{eqnarray}\label{LambdaNC}
\Lambda_{NC}=\left(\frac{12}{\langle \theta^{2} \rangle} \right)^{1/4}=\frac{1}{\lambda} \; ,
\end{eqnarray}
where $\lambda$ is the fundamental length scale that appears in the Klein-Gordon (KG) equation below in (\ref{NCKG}) and in the dispersion relation
(\ref{RelDispDFR}). This approach has the advantage of being unnecessary in order to
specify the form of the function $W$, at least for lowest-order processes.
The study of Lorentz-invariant NC QED,
as Bhabha scattering, dilepton and diphoton production
to LEP data led the authors of \cite{Conroy2003,Carone} to determine the bound
\begin{eqnarray}\label{boundLambda}
\Lambda_{NC} > 160 \; \mbox{GeV} \; \; 95 \% \; \mbox{C.L.} \; .
\end{eqnarray}
The first element of the algebra (\ref{algebraMP}) that commutes with all the others generators $\hat{P}^{\mu}$
and $\hat{M}^{\mu\nu}$ is given by
\begin{eqnarray}
\hat{C}_{1}=\hat{P}_{\mu}\hat{P}^{\mu}-\frac{\lambda^{2}}{2} \, \hat{K}_{ij}\hat{K}_{ij} \; .
\end{eqnarray}
This is the first Casimir operator of the algebra (\ref{algebraMP}). Using the coordinate representation, the operators
$\hat{P}^{\mu}$ and $\hat{K}^{ij}$ can be written in terms of the derivatives
\begin{eqnarray}\label{repcoordinateppi}
\hat{P}_{\mu} \longmapsto i \, \p_{\mu}
\hspace{0.6cm} \mbox{and} \hspace{0.6cm}
\hat{K}_{ij} \longmapsto i \, \frac{\p}{\p \theta^{ij}}= i \, \p_{ij} \; ,
\end{eqnarray}
and consequently, the first Casimir operator in the on-shell condition leads
us to the KG equation in DFR space concerning the scalar field $\phi$
\begin{eqnarray}\label{NCKG}
\left(\Box +\lambda^2\Box_\theta+m^2\right)\phi(x,\theta)=0 \; ,
\end{eqnarray}
where we have defined the D'Alembertian $\theta$-operator
\begin{eqnarray}
\Box_{\theta}=\frac{1}{2} \, \p_{ij} \, \p_{ij}=\frac{1}{2} \, \epsilon_{ijk}\partial_{\theta k} \, \epsilon_{ijl}\partial_{\theta l}
=\partial_{\theta k} \, \partial_{\theta k}=\overrightarrow{\nabla}_{\!\theta}^{\, 2} \; ,
\end{eqnarray}
and $\overrightarrow{\nabla}_{\theta}$ means the $\theta$-gradient operator. The plane wave general solution
for the DFR KG equation is the Fourier integral
\begin{eqnarray}\label{phiXtheta}
\phi( \, x \, , \, \vec{\theta} \, )=\int \frac{d^{4}p}{(2\pi)^{4}} \frac{d^{3}\vec{{\bf k}}}{(2\pi\lambda^{-2})^{3}}
\; \widetilde{\phi}(\, p \, , \, \vec{{\bf k}} \, ) \;  e^{i \, \left( \, p_{\mu} \, x^{\mu}+ \vec{{\bf k}} \, \cdot \, \vec{\theta} \, \right)} \; .
\end{eqnarray}
%
The length $\lambda^{-2}$ is introduced
conveniently in the $k$-integration to maintain the field dimension
as being length inverse. Consequently, the $k$-integration keeps dimensionless.
Substituting the wave plane solution (\ref{phiXtheta}), we obtain the invariant mass
\begin{eqnarray}\label{MassInv}
p^{2}-\lambda^{2} \, \vec{{\bf k}}^{\, 2} =m^2 \; ,
\end{eqnarray}
where $\lambda$ is the parameter with length dimension defined before, it
is a Planck-type length.
Thus we obtain the DFR dispersion relation
\begin{eqnarray}\label{RelDispDFR}
\omega(\, \vec{{\bf p}} \, , \, \vec{{\bf k}} \, )=\sqrt{\vec{{\bf p}}^{\,2}
+\lambda^{2}\, \vec{{\bf k}}^{\,2}+m^2} \; .
\end{eqnarray}
%
%
%
It is easy to see that, using the limit $\lambda \rightarrow 0$ in Eqs.
(\ref{NCKG})-(\ref{RelDispDFR}) we can recover the commutative expression
\cite{EMCAbreuMJNeves2012}.
\begin{figure}[h]
\centering
\includegraphics[scale=0.6]{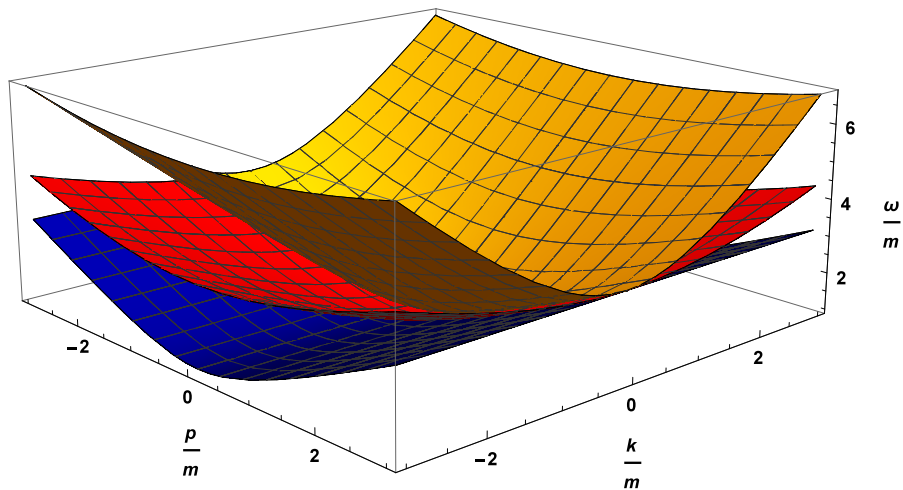}
\caption{The dispersion relation as a function of dimensionless momentum $|\vec{{\bf p}}|/m$,
and of extra-momentum $|\vec{{\bf k}}|/m$, for the values of length
$m \, \lambda=1.0$ (Red surface) and $m \, \lambda=2.0$ (Yellow surface). The commutative case $m \, \lambda=0.0$ is illustrated by the Blue surface.}
\label{GraficoRelDispDFR}
\end{figure}

Since we have constructed the NC KG equation, we will now show its relative action.
We will use the definition of the Moyal-product (\ref{ProductMoyal}) to write the action
for a free complex scalar field in DFR scenario as being
\begin{equation}\label{actionscalarstar}
S_{KG}(\phi^{\ast},\phi)=\int d^{4}x \,d^{3}\theta \, W(\theta) \left( \phantom{\frac{1}{2}} \!\!\!\! \partial_{\mu}\phi^{\ast} \star \partial^{\mu}\phi
+ \lambda^2 \, \partial_{\theta i}\phi^{\ast} \star \partial_{\theta i}\phi
-m^2 \, \phi^{\ast} \star \phi \right) \; ,
\end{equation}
and using the identity (\ref{Idmoyalproduct2}), this free action
can be reduced to the usual one
\begin{eqnarray}\label{actionscalar}
S_{KG}(\phi^{\ast},\phi)=\int d^{4}x \,d^{3}\theta \, W(\theta) \left( \phantom{\frac{1}{2}} \!\!\!\! \left|\partial_{\mu}\phi\right|^{2} +
\lambda^2 \, |\overrightarrow{\nabla}_\theta\phi|^{2}-m^2 \left|\phi\right|^{2} \right) \; ,
\end{eqnarray}
\ni where all the products are the usual ones.

It is easy to show that the DFR Dirac equation can be deduced from the square root of the DFR KG equation,
so we can write the field equation \cite{Amorim5}
\begin{eqnarray}\label{EqDiracDFR}
\Big(\,i \, \slash{\!\!\!\partial}
+\,{{i\lambda}\over2}\,\Gamma_{ij} \, \partial_{ij}
-m \, {\uma} \, \Big) \, \Psi(x,\theta)=0 \; ,
\end{eqnarray}
where $\slash{\!\!\!\partial}:=\gamma^{\mu}\partial_{\mu}$, and $\gamma^{\mu}$'s are the ordinary Dirac matrices that satisfy the usual relation
\begin{eqnarray}\label{gammamu}
\left\{ \gamma^{\mu},\gamma^{\nu} \right\}=2\,\eta^{\mu\nu} \; .
\end{eqnarray}
The matrices $\Gamma^{ij}$ are three matrices $4\times4$ which, by construction, they must be anti-symmetric, i.e., $\Gamma^{ij}=-\Gamma^{ij}$.
We can write the matrices $\Gamma^{ij}$ in terms of Dirac matrices-$\gamma$ by means of the commutation relation
\begin{eqnarray}
\Gamma_{ij}:=\frac{i}{2}\left[\, \gamma_{i} \, , \, \gamma_{j} \,\right]=\sigma_{ij}=\epsilon_{ijk} \, \Sigma_{k} \; ,
\end{eqnarray}
where we can show that the anti-commutation relations are given by
\begin{eqnarray}\label{gammamunu}
\left\{\gamma^{0},\Gamma^{jk} \right\}&=&0 \; ,
\nonumber \\
\left\{\gamma^{i},\Gamma^{jk} \right\}&=&i \, \delta^{i k} \, \gamma^{j}-i \, \delta^{i j} \, \gamma^{k}
+\Gamma^{i j} \, \gamma^{k}-\Gamma^{i k} \, \gamma^{j} \; ,
\nonumber \\
\left \{ \Gamma^{ij},\Gamma^{kl} \right\}&=& \gamma^{k}\, \gamma^{i} \, \delta^{kl}+\gamma^{l} \, \gamma^{j} \, \delta^{ik}
-\gamma^{\lambda} \, \gamma^{i} \, \delta^{jk}
-\gamma^{k}\, \gamma^{j} \,\delta^{il} \; ,
\hspace{1.5cm}
\end{eqnarray}
and the hermiticity property of $\Gamma_{ij}$ is the same as $\gamma^{\mu}$, i.e., $\left(\Gamma_{ij}\right)^{\dagger}=\gamma^{0} \, \Gamma_{ij} \, \gamma^{0}$. This explicit representation
of the matrices $\Gamma_{ij}$ is not a known result in the DFR literature.   If we use these relations,
the Dirac equation (\ref{EqDiracDFR}) leads us to the DFR Klein-Gordon equation.
The components of $\Gamma_{ij}=\epsilon_{ijk} \, \Sigma_{k}$ can be written in terms of the Pauli matrices as
\begin{eqnarray}
%
%
%
\Sigma_{i}=\left(
\begin{array}{cc}
\sigma_{i} & 0 \\
0 & \sigma_{i} \\
\end{array}
\right) \; ,
\end{eqnarray}
which complements the results obtained in \cite{Amorim5}.

It can be shown that the connection between the Dirac equation and its adjoint equation can lead us to a conservation law
\begin{equation}\label{conservlei}
\partial_{\mu}J^{\mu}+\lambda \, \overrightarrow{\nabla}_{\theta}\cdot \vec{J}_{\theta} =0 \; ,
\end{equation}
where $J^{\mu}:=\bar{\Psi}\, \gamma^{\mu} \, \star \, \Psi$, and $J_{\theta i}:=\bar{\Psi}\, \Sigma_{i} \, \star \Psi$
are the currents that emerge from the DFR Dirac equation.
By integrating the expression (\ref{conservlei}) considering the whole space ($x+\theta$), the Dirac field charge
\begin{eqnarray}
Q=\int d^{3}\vec{{\bf r}} \, d^{3}\vec{\theta} \, \, \Psi^{\dagger}(x,\theta) \, \Psi(x,\theta) \; ,
\end{eqnarray}
is conserved, as the commutative usual case.  Notice that the new current term $\bar{\Psi}\, \Sigma_{k} \, \star \Psi$ has the generator of the
rotational group attached to it. In the next section we will investigate the coupling of this current with the gauge fields, which can be an interesting analysis of the Yang-Mills model in DFR phase-space.
The DFR action for the Dirac fermion is
\begin{equation}\label{actionDiracDFRstar}
S_{Dirac}(\bar{\Psi},\Psi)=\int d^{4}x \, d^{3}\theta\, W(\theta) \, \bar{\Psi}(x,\theta) \star
\Big( \, \phantom{\frac{1}{2}} \!\!\!\! \, i \, \slash{\!\!\!\partial}
+\, i \, \lambda \, \overrightarrow{\Sigma} \cdot \overrightarrow{\nabla}_{\theta}
-m \, {\uma} \, \Big) \Psi(x,\theta) \, \,,
\end{equation}
which, using the identity (\ref{ProductMoyal}) can be reduced to
\begin{equation}\label{actionDiracDFRstar2}
S_{Dirac}(\bar{\Psi},\Psi)=\int d^{4}x\,d^{3}\theta\, W(\theta) \, \bar{\Psi}(x,\theta)
\Big(\, \phantom{\frac{1}{2}} \!\!\!\! \, i \, \slash{\!\!\!\partial}
+\, i \, \lambda \, \overrightarrow{\Sigma} \cdot \overrightarrow{\nabla}_{\theta}
-m \, {\uma} \, \Big) \Psi(x,\theta) \, \,,
\end{equation}
which is invariant under symmetry transformations of the Poincar\'e DFR algebra \cite{Amorim5}.
In the next section we will discuss the gauge symmetries of the DFR Dirac Lagrangian.

%

\section{The NC Yang-Mills symmetry revisited}

Let us briefly review the NC Yang-Mills model in DFR framework \cite{YMAbreuNeves2015}, where we have analyzed  the
gauge invariance of the fermion action under the star gauge symmetry transformations. The fermion Lagrangian coupled to
NC gauge fields is given by
\begin{eqnarray}\label{LDiracDcov}
{\cal L}_{fermion-Gauge}=\bar{\Psi}(x,\theta) \star \Big( \, i \, \, \slash{\!\!\!\!D}\star
+\, i \, \overrightarrow{\Sigma} \cdot \overrightarrow{D}_{\theta}\star- m \, {\uma} \, \Big) \Psi(x,\theta)  \; .
\end{eqnarray}
%
%
Here we have defined the NC covariant derivative as being
\begin{eqnarray}\label{DmuA}
D_{\mu}\star = \partial_{\mu}+i \, g \, A_{\mu}\star \; ,
\end{eqnarray}
and $D_{\theta \, i}\star$ is a new antisymmetric star-covariant
derivative associated with the $\theta$-space
\begin{eqnarray}\label{Dmunu}
D_{\theta \, i}\star:=\lambda \, \partial_{\theta \, i} + i \,  g^{\prime} \, A'_{i} \, \star \; ,
\end{eqnarray}
where the field $A'_{i}$ is a vector of three components needed to keep the gauge invariance of the fermion action.
The Lagrangian (\ref{LDiracDcov}) is so manifestly invariant under star-gauge transformations
\begin{eqnarray}\label{PsiABgaugetransf}
\Psi(x,\theta) \hspace{0.1cm} \longmapsto \hspace{0.1cm}
\Psi^{\prime}(x,\theta) \! &=& \! U(x,\theta) \star \Psi(x,\theta) \; ,
\nonumber \\
A_{\mu}(x,\theta) \hspace{0.1cm} \longmapsto \hspace{0.1cm}
A_{\mu}^{\prime}(x,\theta) \! &=& \! U \star A_{\mu}(x,\theta) \star U^{\dagger}
-\frac{i}{g}\, \left(\partial_{\mu}U\right) \star U^{\dagger} \; ,
\nonumber \\
\left(A'_{i}\right)'(x,\theta) \hspace{0.1cm} \longmapsto \hspace{0.1cm}  \left(A'_{i}\right)^{\prime}(x,\theta) \! &=& \! U \star A'_{i}(x,\theta) \star U^{\dagger}
-\frac{i}{g^{\prime}} \, \left(\lambda \, \partial_{\theta \, i} \, U \right) \star U^{\dagger} \; ,
\hspace{0.6cm}
\end{eqnarray}
since we have imposed that the element $U$ is an unitary-star, that is, $U^{\dagger}\star U=\uma$.
The Moyal product of two unitary matrix fields is always unitary, but in general
$\mbox{det}(U \, \star \, U^{\dagger}) \neq \mbox{det}(U) \, \star \, \mbox{det}(U^{\dagger})$, i.e., $\det U \neq 1$.
Therefore, the group that represents the previous star gauge is unitary although not special, say $U^{\star}(N)$. The structure
of $U^{\star}(N)$ is the composition $U^{\star}(N)=U_{N}^{\star}(1) \times SU^{\star}(N)$ of a NC Abelian group with another
NC special unitary group. Concerning gauge symmetry (\ref{PsiABgaugetransf}), we have obtained two NC gauge sectors: the first with a vector gauge
field, and the second one with tensor gauge field. Hence, this gauge symmetry is a composition of two unitary groups, say
$U^{\star}(N)_{A^{\mu}} \times U^{\star}(N)_{A'_{i}}$, and $U$ is the element of both groups.
In fact, the explicit composition of the symmetry group is
\begin{eqnarray*}
U^{\star}(N)_{A^{\mu}} \times U^{\star}(N)_{A_{i}}=U^{\star}(1)_{A^{0\mu}} \times SU^{\star}(N)_{A^{\mu a}} \times U^{\star}(1)_{A_{i}^{' \, 0}} \times SU^{\star}(N)_{A_{i}^{' \, a}} \; .
\end{eqnarray*}
The gauge fields $\left( \, A_{\mu}\, , \, A'_{i} \, \right)$ are hermitian and they can be expanded in terms of the Lie algebra
generators in the adjoint representation as
\begin{eqnarray}
A_{\mu}=A_{\mu}^{0} \, {\uma_{N}} +A_{\mu}^{a} \, G^{a}
\hspace{0.5cm} \mbox{and} \hspace{0.5cm}
A'_{i}  = A_{i}^{' \, 0} \, {\uma_{N}} + A_{i}^{' \, a} \, G^{a}
\end{eqnarray}
where, by satisfying the Lie algebra commutation
relation, we have that
\begin{eqnarray}
\left[\, G^{a} \, , \, G^{b} \, \right]=i \, f^{abc} \, G^{c}
\hspace{0.5cm}
(a,b,c=1,\cdots, N^2-1) \; .
\end{eqnarray}
The fields $A_{\mu}^{0}$ and $A_{i}^{' \, 0}$
come from the Abelian parts $U^{\star}(1)_{A^{0\mu}}$ and $U^{\star}(1)_{\overrightarrow{B}^{0}}$, while the components $A_{\mu}^{a}$ and $\overrightarrow{B}^{a}$ are attached to non-Abelian parts $SU^{\star}(N)_{A^{a\mu}}$ and $SU^{\star}(N)_{\overrightarrow{B}^{a}}$, respectively. The fermion field $\psi$ is the column
matrix of components $\Psi=\left(\psi_{1},\psi_{2}, \cdot \cdot \cdot, \psi_{N}\right)$ which lives in the fundamental
representation of the Lie algebra.

The dynamics in the gauge sector is introduced by the star commutators
\begin{eqnarray}\label{DmuDnuFmunu}
F_{\mu\nu}=-\frac{i}{g} \, \left[ \, D_{\mu} \, , \, D_{\nu} \, \right]_{\star}
=\partial_{\mu}A_{\nu}-\partial_{\nu}A_{\mu}+i \, g \, \left[ \, A_{\mu} \, , \, A_{\nu} \, \right]_{\star}  \; ,
\end{eqnarray}
and
\begin{eqnarray}\label{DmunuDmunuG}
G_{ij}=-\frac{i}{g^{\prime}} \, \left[ \, D_{\theta \, i} \, , \, D_{\theta \, j} \, \right]_{\star}
=\lambda \, \partial_{\theta \, i}A'_{j}
-\lambda \, \partial_{\theta \, j}A'_{i}
+i \, g^{\prime} \, \left[ \, A'_{i} \, , \, A'_{j} \, \right]_{\star} \; .
\end{eqnarray}
By construction, they have the gauge transformations
\begin{eqnarray}\label{transfgaugeFmunu}
F_{\mu\nu} \; \longmapsto \; F_{\mu\nu}^{\;\prime} \! &=& \! U \star F_{\mu\nu} \star U^{\dagger} \; ,
\nonumber \\
G_{ij} \; \longmapsto \;
G^{\;\prime}_{ij} \! &=& \! U \star G_{ij} \star U^{\dagger} \; .
\end{eqnarray}
%
%
Therefore, we have a Lagrangian for the gauge fields given by
\begin{eqnarray}\label{LGauge}
{\cal L}_{Gauge}&=&
-\frac{1}{4}\,\mbox{tr}_{N}\left(F_{\mu\nu}\star F^{\mu\nu}\right)
-\frac{1}{4}\,\mbox{tr}_{N}\left(G_{ij}\star
G_{ij}\right)
\nonumber \\
&&-\frac{1}{2}\,\mbox{tr}_{N}\left(F_{ij} \star G_{ij}
\right) \; .
\end{eqnarray}
The symmetry group is represented by the Lagrangian
\begin{eqnarray}\label{LUNAmuUNBmunu}
\mathcal{L}_{U^{\star}(N)_{A^{\mu}} \times U^{\star}(N)_{B_{i}}}=\mathcal{L}_{Spinor}+\mathcal{L}_{Gauge} \; ,
\end{eqnarray}
which shows independent sectors in DFR formalism.

As an example,  the DFR version of a NC quantum electrodynamics (QED) is represented when $N=1$.
In this case, the non-Abelian subgroup does not contribute for the symmetry,
so we make $G^{a}=0$, $g=e \, Q_{em}$, $g^{\prime}=e^{\prime} \, Q_{em}$.
The gauge field $A^{\mu}$, and the extra field $B_{i}$ have, as generator, the identity matrix,
and the symmetry is composite by
\begin{eqnarray*}
U^{\star}(1)_{em}=U^{\star}(1)_{A^{\mu}} \times U^{\star}(1)_{B^{i}} \; .
\end{eqnarray*}
So the element of each group $U^{\star}(1)$ is given by
\begin{eqnarray}
V(x,\theta)=e^{i \, {\uma} \, \alpha(x,\theta)} \; ,
\end{eqnarray}
and the symmetry is ruled by the transformations
\begin{eqnarray}\label{Localtransfpsi}
\Psi \, \, \longmapsto \, \,
\Psi^{\prime}(x,\theta) &=& e^{i \, \alpha(x,\theta)} \star \Psi(x,\theta) \; ,
\nonumber \\
A_{\mu} \,\, \longmapsto \,\, A_{\mu}^{\prime} &=& V\star A_{\mu}\star V^{\dagger}+\frac{i}{e}\left(\partial_{\mu}V\right) \star V^{\dagger} \; ,
\nonumber \\
A'_{i} \,\, \longmapsto \,\, \left(A'_{i}\right)^{\prime} &=& V \star A'_{i} \star V^{\dagger}+\frac{i}{e^{\prime}}
\left(\lambda\partial_{\theta i}V\right) \star V^{\dagger} \; ,
\end{eqnarray}
where $\Psi$ is a NC spinor of four components, and we have made $Q_{em}=-1$.
%
%
%
The expression of the electromagnetic tensor, and the field strength of $A'_{i}$ are given by
%
\begin{eqnarray}\label{ExpFem}
F_{\mu\nu}&=&\partial_{\mu}A_{\nu}-\partial_{\nu}A_{\mu}
+ i \, e \, \left[ \, A_{\mu} \, , \, A_{\nu} \, \right]_{\star} \; \; \; ,
\nonumber \\
G_{ij}&=&\lambda \, \partial_{\theta i}A'_{j}
-\lambda \, \partial_{\theta j}A'_{i}
+ i \, e^{\prime} \, \left[ \, A'_{i} \, , \, A'_{j} \, \right]_{\star} \;  .
\end{eqnarray}
%
%
%
%
Thereby the NC Abelian version of the Lagrangian (\ref{LGauge}) is simplified as
\begin{eqnarray}\label{LSpinor+gauge}
{\cal L}_{em^{\star}}&=&-\frac{1}{4}\,F_{\mu\nu} \, \star \, F^{\mu\nu}
-\frac{1}{4} \, G_{ij} \, \star \, G_{ij}
-\frac{1}{2} \, F_{ij} \, \star \, G_{ij} \; .
\end{eqnarray}
In the sector of the gauge fields we have naturally the Maxwell Lagrangian in the context of DFR NC.
We can define the components of field strength tensors as
\begin{eqnarray}
F^{\mu\nu}&=&\left(\, {\bf E}^{i} \, , \, \epsilon^{ijk} \, {\bf B}^{k} \, \right)
\nonumber \\
G_{ij}&=&\epsilon_{ijk} \, H_{k} \; ,
\end{eqnarray}
where $H_{i}$ is the dual field of $G_{ij}$, and the Lagrangian more explicitly in terms of these components is
\begin{eqnarray}\label{LEBH}
{\cal L}_{em^{\star}}&=&\frac{1}{2} \left( {\bf E} \star {\bf E}-{\bf B} \star {\bf B}\right)
-\frac{1}{2} \, {\bf H} \star {\bf H}
- {\bf B} \star {\bf H} \; .
\end{eqnarray}

Using the principle of the minimal action associated with respect to $A^{\mu}$, the Lagrangian gives us the NC Maxwell's equations
in the presence of a source $J^{\mu}=\left(\rho,{\bf J}\right)$
\begin{eqnarray}\label{EqfieldA0}
\nabla_{\mu} \, \star \, F^{\,\mu j}
+\epsilon^{\, ijk} \, \, \nabla_{i} \, \star \, H^{\, k} &=& e \, J^{\, j}
\nonumber \\
\nabla_{i} \, \star \, F^{\,i0} &=& e \, \rho \; ,
\end{eqnarray}
where the covariant derivative $\nabla_{\mu}$ acting on strength field tensors is defined by
\begin{eqnarray}\label{nabla0}
\nabla_{\mu} \star F^{\, \mu\nu}:=\partial_{\mu}F^{\, \mu\nu}-e \, \left[A_{\, \mu},F^{\, \mu\nu}\right]_{\star} \; .
\end{eqnarray}
The tensor $F^{\mu\nu}$ must obey the Bianchi identity
\begin{eqnarray}\label{IDBianchi0}
\nabla_{\mu} \star F_{\nu\rho}+\nabla_{\nu} \star F_{\rho\mu}+\nabla_{\rho} \star F_{\mu\nu}=0 \; ,
\end{eqnarray}
which completes the equations for the NC electromagnetism. The field equations for the field $A'_{i}$
are auxiliary equations that emerge exclusively from the NC $\theta$ extra-dimensions
\begin{eqnarray}\label{EqfieldB0}
\epsilon^{ijk} \, \nabla_{\theta}^{\,\, j} \, \star \, H^{k}
+\epsilon^{ijk} \, \nabla_{\theta}^{\,\, j} \, \star \, B^{k}
=-\, e^{\prime} \, \tilde{J}^{\, i} \; ,
\end{eqnarray}
where the anti-symmetric covariant derivative $\nabla_{ij}$ is defined by
\begin{eqnarray}\label{DmunuG}
\epsilon^{ijk} \, \nabla_{\theta}^{\,\, j} \, \star \, H^{k}:=\lambda \, \epsilon^{ijk} \, \partial_{\theta}^{\, j} H^{\, k}
-e^{\prime} \, \, \epsilon^{\, ijk} \,  \left[\, \tilde{B}^{\, j} \, , \, H^{\, k} \,\right]_{\star} \; .
\end{eqnarray}
The current $J^{\,\mu}$ is the classical source of the NC electric and magnetic fields, while the
the anti-symmetric current ${\cal J}^{\, ij}=\epsilon^{ijk} \tilde{J}^{k}$ is interpreted as the source of the background field $H^{k}$
due to the extra-dimension of the NC space-time. When the NC parameter goes to zero, the new current ${\cal J}^{\, ij}$
is automatically zero. From Eqs. (\ref{EqfieldA0}) and (\ref{EqfieldB0}),
we can obtain the conservation law
\begin{eqnarray}\label{EqContinuidade0}
\nabla_{\mu} \star J^{\, \mu} + \nabla_{\theta}^{\,\, k}\star \tilde{J}^{\, k}=0 \; ,
\end{eqnarray}
which expresses the electric charge covariant conservation. The equation (\ref{EqfieldA0})
sets the Ampère Law of the NC magnetic field $B^{k}$, and we have obtained a second Amp\`ere Law
for the background field $H^{k}$. Therefore it can be interpreted as a kind of background magnetic field
emerged of the $\theta$-space. The gauge symmetry permits us to fix a gauge of Coulomb for the $A'_{i}$ field
as
\begin{eqnarray}
\partial_{\theta \, i} A'_{i}=0 \; .
\end{eqnarray}

Back to the Lagrangian in (\ref{LSpinor+gauge}), we will analyze the interactions that arise due to the NC space.
%
%
%
%
Thus the interactions between NC photons and the field $A'_{i}$ have vertex of three and four lines.
The vertex of three lines are
%
%
%
\begin{figure}[!h]
\begin{center}
\newpsobject{showgrid}{psgrid}{subgriddiv=1,griddots=10,gridlabels=6pt}
\begin{pspicture}(0,1.6)(10,3.2)
\psset{arrowsize=0.2 2}
\psset{unit=1}
%
%
%
\pscoil[coilarm=0,coilaspect=0,coilwidth=0.2,coilheight=1.0,linecolor=black](-1.5,1)(1.5,1)
\pscoil[coilarm=0,coilaspect=0,coilwidth=0.2,coilheight=1.0,linecolor=black](-0.1,1.1)(-0.1,3.1)
\put(0.1,3){\large$\gamma$}
\put(-1.6,1.3){\large$\gamma$}
\put(1.3,1.3){\large$\gamma$}
%
%
\pscoil[coilarm=0,coilaspect=0,coilwidth=0.2,coilheight=1.0,linecolor=black](3,1)(5.9,1)
\pscoil[coilarm=0,coilwidth=0.2,coilheight=1.0,linecolor=black](4.4,1.1)(4.4,3.1)
\put(4.7,2.9){\large$A'$}
\put(3,1.3){\large$\gamma$}
\put(5.6,1.3){\large$\gamma$}
%
%
\pscoil[coilarm=0,coilwidth=0.2,coilheight=1.0,linecolor=black](7.5,1)(10.4,1)
\pscoil[coilarm=0,coilaspect=0,coilwidth=0.2,coilheight=1.0,linecolor=black](8.9,1.1)(8.9,3.1)
\put(9.2,2.9){\large$\gamma$}
\put(7.5,1.2){\large$A'$}
\put(10.1,1.2){\large$A'$}
%
%
\end{pspicture}
%
%
%
\end{center}
\end{figure}

\noindent
in which the correspondent Lagrangian is
\begin{eqnarray}\label{Lintvertex3}
{\cal L}_{int}^{\,\,(3)}=- e \, \partial_{\mu}A_{\nu} \, \left[ \, A^{\mu} \, , \, A^{\nu} \, \right]_{\star}
- \lambda \, e \, \partial_{\theta i}A'_{j} \, \left[ \, A_{i} \, , \, A_{j} \, \right]_{\star}
-e^{\prime} \, \partial_{i}A_{j} \, \left[ \, A'_{i} \, , \, A'_{j} \, \right]_{\star} \; .
\end{eqnarray}
For the vertex of four lines we have the diagrams
%
%
\begin{figure}[!h]
\begin{center}
\newpsobject{showgrid}{psgrid}{subgriddiv=1,griddots=10,gridlabels=6pt}
\begin{pspicture}(0,1.3)(10,3.1)
\psset{arrowsize=0.2 2}
\psset{unit=1}
%
%
%
\pscoil[coilarm=0,coilaspect=0,coilwidth=0.2,coilheight=1.0,linecolor=black](-1.5,1)(1.5,1)
\pscoil[coilarm=0,coilaspect=0,coilwidth=0.2,coilheight=1.0,linecolor=black](-0.1,1.1)(-0.1,3.1)
\put(0.1,3){\large$\gamma$}
\put(-1.6,1.3){\large$\gamma$}
\put(1.3,1.3){\large$\gamma$}
%
%
\pscoil[coilarm=0,coilaspect=0,coilwidth=0.2,coilheight=1.0,linecolor=black](3,1)(5.9,1)
\pscoil[coilarm=0,coilwidth=0.2,coilheight=1.0,linecolor=black](4.4,1.1)(4.4,3.1)
\put(4.7,2.9){\large$A'$}
\put(3,1.3){\large$\gamma$}
\put(5.6,1.3){\large$\gamma$}
%
%
\pscoil[coilarm=0,coilwidth=0.2,coilheight=1.0,linecolor=black](7.5,1)(10.4,1)
\pscoil[coilarm=0,coilaspect=0,coilwidth=0.2,coilheight=1.0,linecolor=black](8.9,1.1)(8.9,3.1)
\put(9.2,2.9){\large$\gamma$}
\put(7.5,1.2){\large$A'$}
\put(10.1,1.2){\large$A'$}
%
%
\end{pspicture}
%
%
%
\end{center}
\end{figure}

\noindent
and the Lagrangian is
\begin{equation}
{\cal L}_{int}^{\,\,(4)}=\frac{e^{2}}{4} \, \left[ \, A_{\mu} \, , \, A_{\nu} \, \right]_{\star}^{\; 2}
+\frac{\left(e'\right)^{2}}{4} \, \left[ \, A'_{i} \, , \, A'_{j} \, \right]_{\star}^{\; 2}
+\frac{e \, e^{\prime}}{2} \, \left[ \, A_{i} \, , \, A_{j} \, \right]_{\star}
\, \left[ \, A'_{i} \, , \, A'_{j} \, \right]_{\star} \; ,
\end{equation}
where we have achieved an invariance property for the DFR Dirac action under the local transformations
(\ref{PsiABgaugetransf}). To guarantee this invariance we must introduce an extra vector field $A'_{i}$ in (\ref{Dmunu}), in addition to the
usual vector field $A^{\mu}$.
The field $A^{\mu}$ has the first transformation of (\ref{ABgaugetransf}), while $A'_{i}$ has the second transformation of (\ref{ABgaugetransf}).
This new vector field is associated with the $\theta$-space and it must to be attached to those extra-dimensions.
The NC gauge theory obtained here recovers the standard case of $SU(N)$ Yang-Mills,
and the usual $U(1)$ QED, in the commutative limit $\theta=0$ and taking $\lambda=0$ in the Lagrangian (\ref{LDiracDcov}) and
(\ref{LGauge}).

In the next section, based on this gauge symmetry group, we will propose a NC electroweak model in DFR framework via the inclusion of
new antisymmetrical bosons.

\pagebreak

\section{The Model $U_{L}^{\star}(2)_{A^{\mu}} \times U_{R}^{\star}(1)_{X^{\mu}} \times U_{L}^{\star}(2)_{B_{i}} \times U_{R}^{\star}(1)_{C_{i}}$}
\renewcommand{\theequation}{3.\arabic{equation}}
\setcounter{equation}{0}

The composite model $U_{L}^{\star}(2)_{A^{\mu}} \times U_{R}^{\star}(1)_{X^{\mu}} \times U_{L}^{\star}(2)_{B_{i}} \times U_{R}^{\star}(1)_{C_{i}}$ is the DFR version of the GSW model for electroweak interactions. We note that the structure $U_{L}^{\star}(2)_{A^{\mu}} \times U_{R}^{\star}(1)_{X^{\mu}}$,
contained in the complete symmetry is the electroweak model in the NC space-time with $\theta^{\mu\nu}$ parameter constant
\cite{Chaichian2003}. We need the structure of this extended NC group in order to introduce the left and right-handed fermion sectors.
We will associate the left sector with the groups $U_{L}^{\star}(2)_{A^{\mu}}$ and $U_{L}^{\star}(2)_{B_{i}}$, while the right sector
is attached to the groups $U_{R}^{\star}(1)_{X^{\mu}}$ and $U_{R}^{\star}(1)_{C_{i}}$. The groups
$U_{L}^{\star}(2)$ are formed by two subgroups, i.e., $U_{L}^{\star}(2)=U^{\star}(1) \times SU_{L}^{\star}(2)$, where $U^{\star}(1)$ is a NC Abelian, and
$SU_{L}^{\star}(2)$ is a NC non-Abelian subgroup. We denote the first one by $U_{L}^{\star}(2)_{A^{\mu}}=U^{\star}(1)_{A^{0\mu}} \times SU_{L}^{\star}(2)_{A^{\mu a}}$, and the second one by $U_{L}^{\star}(2)_{B_{i}}=U^{\star}(1)_{B_{i}^{\, 0}} \times SU_{L}^{\star}(2)_{B_{i}^{\,a}}$.

As in the usual model, we can introduce the projectors $P_{L}=\left(\openone-\gamma^{5}\right)/2$
and $P_{R}=\left(\openone+\gamma^{5}\right)/2$, where $\Psi_{L}=P_{L}\,\Psi$, and $\Psi_{R}=P_{R}\,\Psi$. The projectors satisfy the properties
\begin{eqnarray}\label{PLPR}
P_{L}^{2}=P_{L}
\hspace{0.2cm} , \hspace{0.2cm}
P_{R}^{2}=P_{R}
\hspace{0.2cm} , \hspace{0.2cm}
P_{L} \, P_{R}=P_{R} \, P_{L}=0
\hspace{0.2cm} , \hspace{0.2cm}
P_{L}+P_{R}=\openone \; \; ,
\end{eqnarray}
and with the matrices $\left(\gamma^{\mu},\Gamma^{\mu\nu}\right)$ of the NC Dirac equation we have that
\begin{eqnarray}\label{PLPRGammamu}
P_{L} \, \gamma^{\mu}=\gamma^{\mu} \, P_{R}
\hspace{0.2cm} , \hspace{0.2cm}
P_{R} \, \gamma^{\mu}=\gamma^{\mu} \, P_{L}
\hspace{0.2cm} , \hspace{0.2cm}
P_{L} \, \Sigma^{i}=\Sigma^{i} \, P_{L}
\hspace{0.2cm} , \hspace{0.2cm}
P_{R} \, \Sigma^{i}=\Sigma^{i} \, P_{R} \; ,
\end{eqnarray}
where $\{\, \gamma^{5} \, , \, \gamma^{\mu} \, \}=\left[ \, \gamma^{5} \, , \, \Sigma^{i} \, \right]=0$, and
$\gamma^{5}=i \, \gamma^{0}\gamma^{1}\gamma^{2}\gamma^{3}$.
Thus we can construct the gauge transformations that keep invariant the Lagrangian of the model
$U_{L}^{\star}(2)_{A^{\mu}} \times U_{R}^{\star}(1)_{X^{\mu}} \times U_{L}^{\star}(2)_{B_{i}} \times U_{R}^{\star}(1)_{C_{i}}$.
Concerning the groups $U_{L}^{\star}(2)_{A^{\mu}}$ and $U_{L}^{\star}(2)_{B_{i}}$,
we define the fermion doublet (neutrinos and leptons left-handed) that turns in the fundamental representation of groups $U_{L}^{\star}(2)$,
and the same happens with the anti-fundamental representation of groups $U_{R}^{\star}(1)$
\begin{eqnarray}
\Psi_{L}(x,\theta)=
\left(
\begin{array}{c}
\nu_{\ell L} \\
\ell_{L} \\
\end{array}
\right)
\longmapsto \Psi_{L}^{\prime}(x,\theta)= U(x,\theta)
\star \Psi_{L}(x,\theta) \star V_{2}^{-1}(x,\theta) \; ,
\end{eqnarray}
where $U(x,\theta)$ is the element of the group $U_{L}^{\star}(2)$, and $V_{2}$ is the element of $U_{R}^{\star}(1)$.
Concerning the right sector $U_{R}^{\star}(1)$, any fermion (left or right) doublet can be transformed into an anti-fundamental representation such that
\begin{eqnarray}
\psi(x,\theta) \longmapsto \psi^{\prime}(x,\theta)= \psi(x,\theta) \star V_{2}^{-1}(x,\theta) \; ,
\end{eqnarray}
where $\psi$ can represent any lepton $\left(\, \Psi_{L} \, , \, \ell_{R} \, \right)$. 
The covariant derivatives acting on fermions into the left and right-sectors of the model can be defined in the following way
\begin{eqnarray}
D_{L \, \mu} \, \Psi_{L} \! &=& \! \partial_{\mu} \, \Psi_{L}+i \, g_{1} \, A_{\mu} \star \Psi_{L}-i \, J_{L} \, g_{1}^{\prime} \, \Psi_{L} \star X_{\mu} \; ,
\nonumber \\
D_{L \, \theta \, i} \, \Psi_{L} \! &=& \! \lambda \, \partial_{\theta \, i} \, \Psi_{L} + i \, g_{2} \, B_{i} \star \Psi_{L}
- i \, J_{L} \, g_{2}^{\prime} \, \Psi_{L} \star C_{i} \; ,
\nonumber \\
D_{R \, \mu} \, \ell_{R}\! &=& \! \partial_{\mu} \, \ell_{R} - i \, J_{R} \, g_{1}^{\prime} \, \ell_{R} \star X_{\mu} \; ,
\nonumber \\
D_{R \, \theta \, i} \, \ell_{R}\! &=& \!\lambda \, \partial_{\theta \, i} \, \ell_{R} - i \, J_{R} \, g_{2}^{\prime} \, \ell_{R} \star C_{i} \; ,
\end{eqnarray}
where
\begin{eqnarray}
A_{\mu}=A_{\mu}^{\,0}\,{\openone_{2}}+A_{\mu}^{a} \, \frac{\sigma^{a}}{2}
\hspace{0.5cm} \mbox{and} \hspace{0.5cm}
B_{i}= B_{i}^{\, 0}\,{\openone_{2}}+B_{i}^{\, a} \, \frac{\sigma^{a}}{2}
\end{eqnarray}
are the NC gauge fields of $U_{L}^{\star}(2)$ groups, and $X^{\mu}$, $C_{i}$ are the NC Abelian
gauge fields of $U_{R}^{\star}(1)$ groups. Here we have used the symbol $J$ as the generator of $U_{R}^{\star}(1)$.
Defining the following gauge transformations
\begin{eqnarray}\label{ABgaugetransf}
A_{\mu}(x,\theta) \,\, \longmapsto \,\, A_{\mu}^{\prime}(x,\theta) \! &=& \! U \star A_{\mu}(x,\theta)\star U^{-1}
-\frac{i}{g_{1}} \, U \star \partial_{\mu}U^{-1} \; ,
\nonumber \\
B_{i}(x,\theta) \,\, \longmapsto \,\, B_{i}^{\prime}(x,\theta) \! &=& \! U \star B_{i}(x,\theta) \star U^{-1}
-\frac{i}{g_{2}} \, U \star \lambda \, \partial_{\theta \, i} U^{-1} \; ,
\nonumber \\
X_{\mu}(x,\theta) \,\, \longmapsto \,\, X_{\mu}^{\prime}(x,\theta) \! &=& \!  V_{2} \star X_{\mu}(x,\theta) \star V_{2}^{-1}
-\frac{i}{J\,g_{1}^{\prime}} \, V_{2} \star \partial_{\mu} V_{2}^{-1} \; ,
\nonumber \\
C_{i}(x,\theta) \,\, \longmapsto \,\, C_{i}^{\prime}(x,\theta) \! &=& \! V_{2} \star C_{i}(x,\theta) \star V_{2}^{-1}
-\frac{i}{J\,g_{2}^{\prime}} \, V_{2} \star \lambda \, \partial_{\theta \,i} V_{2}^{-1} \; ,
\hspace{0.6cm}
\end{eqnarray}
we can obtain that the previous covariant derivatives have the following transformations
\begin{eqnarray}
D_{L \, \mu} \, \Psi_{L} \longmapsto \left(D_{L\mu} \, \Psi_{L}\right)^{\prime} &=& U \star D_{L\mu} \, \Psi_{L} \star V_{2}^{-1} \; ,
\nonumber \\
D_{L \, \theta \, i}\Psi_{L} \longmapsto \left(D_{L \, \theta \, i} \, \Psi_{L}\right)^{\prime} &=& U \star D_{L \, \theta \, i} \, \Psi_{L} \star V_{2}^{-1} \; ,
\nonumber \\
D_{R \, \mu} \, \ell_{R} \longmapsto \left(D_{R \, \mu} \, \ell_{R}\right)^{\prime} &=& D_{R \, \mu} \, \ell_{R} \star V_{2}^{-1} \; ,
\nonumber \\
D_{R \, \theta \, i} \, \ell_{R} \longmapsto \left(D_{R \, \theta \, i} \, \ell_{R}\right)^{\prime} &=& D_{R \, \theta \, i} \, \ell_{R} \star V_{2}^{-1} \; .
\end{eqnarray}
Therefore, the invariant leptons' Lagrangian under the previous gauge transformations is
\begin{equation}\label{Lleptons}
{\cal L}_{Leptons}=\bar{\Psi}_{L}\star \, i \, \slash{\!\!\!\!D}_{L} \star \Psi_{L}
+\bar{\ell}_{R} \star \, i \, \slash{\!\!\!\!D}_{R} \star \ell_{R} \; .
\end{equation}
According to the properties (\ref{PLPR}) and (\ref{PLPRGammamu}), it is easy to see that the terms
of the derivatives $D_{L \, \theta \, i}$ and $D_{R \, \theta \, i}$ that could emerge from (\ref{Lleptons}) are zero, that is,
\begin{eqnarray}\label{PsiLGammamunuPsiL}
\bar{\Psi}_{L} \star \overrightarrow{\Sigma} \cdot \overrightarrow{D}_{L \, \theta} \, \Psi_{L}
=\bar{\ell}_{R} \star \overrightarrow{\Sigma} \cdot \overrightarrow{D}_{R \, \theta} \, \ell_{R}=0 \; .
\end{eqnarray}
Consequently, if we introduce left and right handed components, the gauge invariant terms finishes the propagation in the $\theta$-space,
and also all the fermions interactions have the sector of the anti-symmetrical gauge fields. This problem will be solved when we will introduce
the Yukawa interactions in the Higgs sector in order to break the gauge symmetry shown in this section.

In the gauge fields sector, the field strength tensors of the bosons are defined by
\begin{eqnarray}\label{Fmunu}
F_{\mu\nu} \!&=&\! \partial_{\mu}A_{\nu}-\partial_{\nu}A_{\mu}+i \, g_{1} \, \left[ \, A_{\mu} \, , \, A_{\nu} \, \right]_{\star}  \; ,
\nonumber \\
X_{\mu\nu} \!&=&\! \partial_{\mu}X_{\nu}-\partial_{\nu}X_{\mu}+i \, J \, g_{1}^{\prime} \, \left[ \, X_{\mu} \, , \, X_{\nu} \, \right]_{\star} \; ,
\end{eqnarray}
and
\begin{eqnarray}
G_{ij}  \!&=&\! \lambda \, \partial_{\theta \, i}B_{j}-\lambda \, \partial_{\theta \, j}B_{i}
+ i \, g_{2} \, \left[ \, B_{i} \, , \, B_{j} \, \right]_{\star} \; ,
\nonumber \\
C_{ij} \!&=&\! \lambda\partial_{\theta \, i} \, C_{j}-\lambda \, \partial_{\theta \, j} \, C_{i}+i \, J \, g_{2}^{\prime} \, \left[ \, C_{i} \, , \, C_{j} \, \right]_{\star} \; .
\end{eqnarray}
%
%
We can write it in the basis, $({\openone_{2}},\frac{\sigma^{a}}{2})$ where the components of $F$ and $G$ are given by
\begin{eqnarray}
F_{\mu\nu}^{\,\,0} \!&=&\! \partial_{\mu}A_{\nu}^{\,0}-\partial_{\nu}A_{\mu}^{\,0}
+i \, g_{1} \left[ \, A_{\mu}^{\, 0} \, , \, A_{\nu}^{\, 0} \, \right]_{\star}
+\frac{i}{4} \, g_{1} \, \left[ \, A_{\mu}^{\, a} \, , \, A_{\nu}^{\, a} \, \right]_{\star} \; ,
\nonumber \\
F_{\mu\nu}^{\,\,a} \!&=&\! \partial_{\mu}A_{\nu}^{\,a}-\partial_{\nu}A_{\mu}^{\,a}
-\frac{1}{2} \, g_{1} \, \varepsilon^{abc} \left\{ \, A_{\mu}^{b} \, , \, A_{\nu}^{c} \, \right\}_{\star}
+ \, i \, g_{1} \, \left[ \, A_{\mu}^{\,a} \, , \, A_{\nu}^{\,0} \, \right]_{\star}
\!+i \, g_{1} \, \left[ \, A_{\mu}^{\,0} \, , \, A_{\nu}^{\,a} \, \right]_{\star} \; ,
\hspace{-0.5cm}\nonumber \\
\end{eqnarray}
and
\begin{eqnarray}
G_{ij}^{\,\,0} \!&=&\! \lambda \, \partial_{ \theta \, i}B_{j}^{\,0}-\lambda \, \partial_{ \theta \, j}B_{i}^{\,0}
+ i \, g_{2} \, \left[ \, B_{i}^{\, 0} \, , \, B_{j}^{\, 0} \, \right]_{\star}
+\frac{i}{4} \, g_{2} \, \left[ \, B_{i}^{\, a} \, , \, B_{j}^{\, a} \, \right]_{\star} \; ,
\nonumber \\
G_{ij}^{\,\,a} \!&=&\! \lambda \, \partial_{\theta i}B_{j}^{\,a}-\lambda \, \partial_{ \theta \, j}B_{i}^{\,a}
-\frac{1}{2} \, g_{2} \, \varepsilon^{abc} \left\{ \, B_{\mu\nu}^{b} \, , \, B_{\rho\lambda}^{c} \, \right\}_{\star}
+i \, g_{2} \, \left[ \, B_{i}^{\,a} \, , \, B_{j}^{\,0} \, \right]_{\star}
\!+ i \, g_{2} \, \left[ \, B_{i}^{\,0} \, , \, B_{j}^{\,a} \, \right]_{\star} \; .
\hspace{-0.5cm}\nonumber \\
\end{eqnarray}
The Lagrangian of the gauge fields, invariant under the complete symmetry, is
\begin{eqnarray}\label{Lgauge}
{\cal L}_{Gauge}&=&-\frac{1}{2} \,\mbox{tr}\left(F_{\mu\nu}\star F^{\mu\nu}\right)
-\frac{1}{4} \, X_{\mu\nu} \star X^{\mu\nu}
-\frac{1}{2} \, \mbox{tr}\left(G_{ij} \star G_{ij}\right)
\nonumber \\
&&
-\frac{1}{4} \, C_{ij} \star C_{ij}
- \, \mbox{tr}\left( F_{ij} \star G_{ij} \right)
-\frac{1}{2} \, C_{ij} \star X_{ij} \; .
\end{eqnarray}
%
%
%

The interaction terms that emerge from (\ref{Lleptons}) reveal how the leptons
and neutrinos can interact with the gauge fields components. Initially, we can write these interactions as
\begin{eqnarray}\label{LintAY}
{\cal L}_{Leptons-Gauge}^{\, int}\!\!\!\!
&&=-\frac{g_{1}}{2} \, \, \bar{\nu}_{\ell L} \star \left( \, \slash{\!\!\!\!A}^{1}-i \, \slash{\!\!\!\!A}^{2} \right) \star \ell_{L}
-\frac{g_{1}}{2} \, \, \bar{\ell}_{L} \star \left( \, \slash{\!\!\!\!A}^{1} +i \, \slash{\!\!\!\!A}^{2} \right) \star \nu_{\ell L}
\nonumber \\
&&-\bar{\Psi}_{L} \star \left( \, g_{1} \, \slash{\!\!\!\!A}^{3} \, I^{3}+g_{1} \, \slash{\!\!\!\!A}^{0}- J_{L} \, g_{1}^{\prime} \, \slash{\!\!\!\!X}
\, \right)
\star \Psi_{L}
+\bar{\ell}_{R}\star \left( \, J_{R} \, g_{1}^{\prime} \, \slash{\!\!\!\!X} \, \right)\star \ell_{R}
\nonumber \\
&&+\bar{\Psi}_{L}\star \gamma^{\mu}\left[ \!\!\! \phantom{\frac{1}{2}} \Psi_{L}, J_{L} \, g_{1}^{\prime} \, X_{\mu} \right]_{\star}
+\bar{\ell}_{R}\star \gamma^{\mu}\left[ \!\!\! \phantom{\frac{1}{2}} \ell_{R}, J_{R} \, g_{1}^{\prime} \, X_{\mu}  \right]_{\star}
\; ,
\end{eqnarray}
where we have defined $I^{3}=\sigma^{3}/2$, for simplicity.
This last expression can motivate us to define the charged bosons $W^{\pm}$ in the
usual manner, i.e., $\sqrt{2} \, W_{\mu}^{\pm}=A_{\mu}^{1}\mp i \, A_{\mu}^{2}$, which defines the interaction between leptons/neutrinos and $W^{\pm}$.
In terms of the interaction involving leptons and neutral bosons we can observe the need to define which one is the hypercharge generator
of the model. This task will be done in the next section.

\pagebreak

\section{The first SSB and the hypercharge generator}
\renewcommand{\theequation}{4.\arabic{equation}}
\setcounter{equation}{0}

The proposed model has the structure of four Abelian groups and two non-Abelian groups, namely,
\begin{eqnarray*}
U^{\star}(1)_{A^{\mu 0}} \times SU_{L}^{\star}(2)_{A^{\mu a}} \times U_{R}^{\star}(1)_{X^{\mu}} \times  U^{\star}(1)_{B_{i}^{0}}
\times SU_{L}^{\star}(2)_{B_{i}^{\, a}} \times U_{R}^{\star}(1)_{C_{i}} \; .
\end{eqnarray*}
%
To identify the hypercharge generator, we will introduce the first Higgs sector
to break one of the two Abelian NC symmetries, and consequently, we will eliminate the residual symmetry $U^{\star}(1)_{A^{\mu 0}}$ of the model.
Then we will denote this Higgs field as the Higgs-one $\Phi_{1}$. After this first spontaneous breaking symmetry, we will obtain a remaining symmetry such as
\begin{eqnarray}
U^{\star}(1)_{A^{\mu0}} \times SU_{L}^{\star}(2)_{A^{\mu a}} \times U_{R}^{\star}(1)_{X^{\mu}} \times  U^{\star}(1)_{B_{i}^{\, 0}} \times SU_{L}^{\star}(2)_{B_{i}^{\,a}} \times U_{R}^{\star}(1)_{C_{i}}
\nonumber \\
\stackrel{\langle \Phi_{1} \rangle}{\longmapsto} SU_{L}^{\star}(2)_{A^{\mu a}} \times U_{Y}^{\star}(1) \times U^{\star}(1)_{G_{i}} \times U_{R}^{\star}(1)_{C_{i}} \; .
\end{eqnarray}
The group $U_{Y}^{\star}(1)$ appears from a mixing of the groups $U^{\star}(1)_{A^{\mu 0}}$, and $U_{R}^{\star}(1)_{X^{\mu}}$.
To do that, we will introduce the Higgs-one Lagrangian
\begin{eqnarray}\label{LHiggs1}
{\cal L}_{Higgs}^{\, (1)}={\cal L}_{Scalar}^{\, (1)}+{\cal L}_{Yukawa}^{\, (1)} \; ,
\end{eqnarray}
where we can define a scalar sector
\begin{equation}\label{LScalar1}
{\cal L}_{Scalar}^{\, (1)}=\left(D_{\mu} \, \Phi_{1}\right)^{\dagger} \star D^{\mu} \, \Phi_{1}
+ \left(D_{\theta \, i} \, \Phi_{1}\right)^{\dagger} \star D_{\theta \, i} \, \Phi_{1}
-\mu_{1}^2 \left(\Phi_{1}^{\,\dagger} \star \Phi_{1}\right)-g_{H1} \left(\Phi_{1}^{\,\dagger} \star \Phi_{1}\right)^{2} \; ,
\end{equation}
and the Yukawa sector
\begin{eqnarray}\label{LYukawa1}
{\cal L}_{Yukawa}^{(1)} \!&=&\! \frac{i}{2} \, f_{1} \, \bar{\Psi}_{L}\star \Phi_{1} \star \overrightarrow{\Sigma} \cdot \overrightarrow{D}_{R\,\theta}\ell_{R}
-\frac{i}{2} \, f_{1}^{\ast} \, \overline{\overrightarrow{D}_{R \, \theta} \,\ell}_{R} \cdot \overrightarrow{\Sigma} \star \Phi_{1}^{\dagger} \star \Psi_{L}
\nonumber \\
&&+\frac{i}{2} \, f_{2} \; \bar{\ell}_{R}\star \Phi_{1}^{\dagger} \star \overrightarrow{\Sigma} \cdot \overrightarrow{D}_{L\,\theta} \Psi_{L}
-\frac{i}{2} \, f_{2}^{\ast} \, \overline{\overrightarrow{D}_{L\, \theta}\Psi}_{L} \cdot \overrightarrow{\Sigma} \star \Phi_{1} \star \ell_{R} \; .
\end{eqnarray}
Here, $\mu_{1}$, $g_{H1}$ are real parameters, and $f_{i}\, (i=1,2)$ are complex ones.
The covariant derivatives of (\ref{LScalar1}) are defined in terms of the NC Abelian gauge fields, and they act on the Higgs-one as
\begin{eqnarray}\label{DmuPhi1}
D_{\mu}\,\Phi_{1}\!&=&\!\partial_{\mu} \, \Phi_{1}+i \, g_{1} \, A_{\mu}^{\; 0} \star \Phi_{1}-i \, J_{\Phi_{1}} \, g_{1}^{\prime} \, \Phi_{1} \star X_{\mu} \; ,
\nonumber \\
D_{\theta \, i}\,\Phi_{1}\!&=&\!\lambda \, \partial_{\theta \, i} \, \Phi_{1}+i \, g_{2} \, B_{i}^{\; 0} \star \Phi_{1}
+i \, g_{2} \, B_{i}^{\, a} \, \frac{\sigma^a}{2}\star\Phi_{1} \; ,
\end{eqnarray}
where the sector of $D_{\mu}$ is coupled to the Higgs-one through the NC Abelian fields, although $D_{\theta \, i}$ is coupled to $\Phi_{1}$
via the antisymmetric non-Abelian fields of both $U_{L}^{\star}(2)$. Therefore, the field $\Phi_{1}$ is a scalar doublet of the NC groups
$U_{L}^{\star}(2)$. In the antisymmetric sector, the Higgs-one transforms into the fundamental representation of $U_{L}^{\star}(2)$ as
\begin{eqnarray}\label{transfPhi1U}
\Phi_{1}=
\left(
\begin{array}{c}
\phi_{1}^{ \, (+)} \\
\phi_{1}^{\, (0)} \\
\end{array}
\right)
\longmapsto \Phi_{1}^{\prime}=U \star \Phi_{1} \; ,
\end{eqnarray}
where $U$ is the element of both groups $U_{L}^{\star}(2)$. In the NC Abelian sector, $\Phi_{1}$ transforms in the
fundamental representation of $U^{\star}(1)_{A^{\mu0}}$, and in the anti-fundamental representation of
$U_{R}^{\star}(1)_{X^{\mu}}$ as
\begin{equation}\label{transfPhi1VV}
\Phi_{1} \longmapsto \Phi_{1}^{\; \prime}= V_{1} \star \Phi_{1} \star V_{2}^{-1} \; ,
\end{equation}
where $V_{1}$ is the element of the subgroup $U^{\star}(1)_{A^{\mu0}}$. Using the gauge transformations,
we will obtain that the covariant derivative $D_{\mu}$ transforms like (\ref{transfPhi1VV}), and $D_{\mu\nu}$
transforms like (\ref{transfPhi1U}). It is easy to see that the Higgs potential in (\ref{LScalar1}) is unaltered
by both transformations (\ref{transfPhi1VV}) and (\ref{transfPhi1U}).

The minimal value of the previous Higgs potential can be obtained by the non-trivial vacuum expected value (VEV) of
the Higgs field that keeps the translational invariance in the space $x+\theta$, where we can choose it as the
doublet of the constant VEV
\begin{eqnarray}
\langle \Phi_{1} \rangle_{0}=
\left(
  \begin{array}{c}
    0 \\
    \frac{u}{\sqrt{2}} \\
  \end{array}
\right) \; ,
\end{eqnarray}
where $u$ is the non-trivial vacuum expectation value (VEV) of the Higgs field $\Phi_{1}$, defined by $u:=\sqrt{-\mu_{1}^{2} / g_{H1}}$, when $\mu_{1}^{2}<0$. We choose the parametrization of the $\Phi_{1}$-complex field as
\begin{eqnarray}\label{PhiGaugeunitary}
\Phi_{1}(x,\theta)= \frac{\left(u+H_{1}\right)}{\sqrt{2}} \star e^{i\frac{\sigma^{a}}{2}\frac{\chi^{a}}{u}}
\left(
  \begin{array}{c}
    0 \\
    1 \\
  \end{array}
\right) \; ,
\end{eqnarray}
where $\chi^{a}$ and $H_{1}$ are real functions of $(x,\theta)$. Now under the gauge transformation (\ref{transfPhi1U}),
we will fix $\chi^{a}=-u\, \omega^{a}$ to eliminate the Goldstone bosons (unitary gauge), such that
$e^{i\frac{\sigma^{a}}{2}\frac{\chi^{a}}{u}}\star e^{i\frac{\sigma^{a}}{2}\,\omega^{a}}=\openone$, and the $\Phi_{1}$-field
has the form
\begin{eqnarray}\label{PhiH1}
\Phi_{1}(x,\theta)=
\frac{u+H_{1}(x,\theta)}{\sqrt{2}}
\left(
  \begin{array}{c}
    0 \\
    1 \\
  \end{array}
\right) \; .
\end{eqnarray}
This VEV defines a scale concerning the breaking of the residual symmetry $U^{\star}(1)_{A^{\mu0}}$, where the NC Abelian
gauge field acquires a mass term. So the free part of the Lagrangian is given by
\begin{eqnarray}\label{LHiggs1Masses}
{\cal L}_{Scalar-0}^{(1)}\!\!\!&&=\frac{1}{2}\,\left(\partial_{\mu} \, H_{1}\right)^{2}
+\frac{\lambda^{2}}{2}\left(\overrightarrow{\nabla}_{\theta} \, H_{1}\right)^{2}
-\frac{1}{2}\left(-2\mu_{1}^{2}\right)H_{1}^{2}
+\frac{g_{2}^{2} \, u^{2}}{4} \, B_{i}^{+} \, B_{i}^{\, -}
\nonumber \\
&& +\frac{u^{2}}{2} \left(\phantom{\frac{1}{2}} \!\!\! g_{1} \, A_{\mu}^{0}- J_{\Phi_{1}} \, g_{1}^{\prime} \, X_{\mu} \, \right)^{2}
\!+\frac{u^{2}}{4} \left( \, -\frac{1}{2} \,  g_{2} \, B_{i}^{\, 3} + g_{2} \, B_{i}^{\, 0} \, \right)^{2} \; ,
\end{eqnarray}
where we have used the Moyal product identity (\ref{ProductMoyal}). In this expression we have obtained
the Higgs $H_{1}$-field Lagrangian with propagation along the $\theta$-space, and a mass of $M_{H_{1}}=\sqrt{2 \, g_{1H} \, u^{2}}$.
We observe the emergence of a mass term for a new charged field $B^{\pm}$ defined by the combinations
\begin{eqnarray}\label{B+-}
B_{i}^{\,\pm}&=&\frac{B_{i}^{\, 1}\mp iB_{i}^{\, 2}}{\sqrt{2}} \; ,
\end{eqnarray}
where it mass is
\begin{eqnarray}
m_{B^{\pm}}=\frac{1}{2} \, g_{2} \, u \; .
\end{eqnarray}
To define the weak hypercharge generator, the mass terms in (\ref{LHiggs1Masses}) motivate us to introduce the orthogonal transformations
\begin{eqnarray}\label{transfA0CGY}
A_{\mu}^{\, 0} \!&=&\! \cos\alpha \, G_{\mu}+ \sin\alpha \, Y_{\mu}\,\,,
\nonumber \\
X_{\mu} \!&=&\! -\sin\alpha \, G_{\mu}+\cos\alpha \, Y_{\mu} \; ,
\end{eqnarray}
and
\begin{eqnarray}\label{transfB0X}
B_{i}^{\, 0} \!&=&\! \cos\beta \, G\,'_{i} + \sin\beta \, Y\,'_{i}\,\,,
\nonumber \\
C_{i} \!&=&\! -\sin\beta \, G\,'_{i}+\cos\beta \, Y\,'_{i} \; ,
\end{eqnarray}
where $\alpha$ and $\beta$ are the mixing angles. Here, the fields $Y^{\mu}$ are the gauge fields associated with the hypercharge
generator. For example, we can use the transformation (\ref{transfA0CGY}) into the covariant derivative $D_{\mu}$
acting on Higgs-$\Phi_{1}$ to identify the hypercharge it generator as
\begin{eqnarray}\label{DmuPhi1Y}
D_{\mu} \, \Phi_{1} \!&=&\! \partial_{\mu} \, \Phi_{1}+i \left( \, g_{1} \, \sin\alpha - J_{\Phi_{1}} \, g_{1}^{\prime} \, \cos\alpha \, \right) Y_{\mu} \star \Phi_{1} \, + \cdots \; ,
\end{eqnarray}
where we have defined
\begin{eqnarray}\label{gYphi1}
g \, Y_{\Phi_{1}}= g_{1} \, \sin\alpha - J_{\Phi_{1}} \, g_{1}^{\prime} \, \cos\alpha \; .
\end{eqnarray}
The orthogonal transformation (\ref{transfA0CGY}) allows us to fix the relation
\begin{eqnarray}\label{tanalpha}
\tan\alpha=\frac{g_{1}^{\prime}}{g_{1}} \; ,
\end{eqnarray}
so the expression (\ref{gYphi1}) is given by
\begin{eqnarray}\label{gYphi1}
g \, Y_{\Phi_{1}}= g_{1} \, \sin\alpha - J_{\Phi_{1}} \, g_{1}^{\prime} \, \cos\alpha \; .
\end{eqnarray}
Here this result suggests to define the new coupling constant as
\begin{eqnarray}\label{gg1g1prime}
g=g_{1} \, \sin\alpha = g_{1}^{\, \prime} \, \cos\alpha \; ,
\end{eqnarray}
and the hypercharge of Higgs-$\Phi_{1}$ is given by
\begin{eqnarray}
Y_{\Phi_{1}}=1-J_{\Phi_{1}} \; .
\end{eqnarray}
Since the Higgs-$\Phi_{1}$ is a singlet with relation to subgroup
$U^{\star}(1)_{A^{\mu 0}} \times U_{R}^{\star}(1)_{X^{\mu}}$, we can fix $J_{\Phi_{1}}=1$, so we obtain $Y_{\Phi_{1}}=0$.

Therefore, with these definitions the Lagrangian (\ref{LHiggs1Masses}) can be written as
\begin{eqnarray}\label{LHiggs1Masses2}
{\cal L}_{Scalar-0}^{(1)}\!\!\!&&=\frac{1}{2} \, \left(\partial_{\mu} \, H_{1}\right)^{2}
+\frac{\lambda^{2}}{2} \, \left( \overrightarrow{\nabla}_{\theta}\,  H_{1}\right)^{2}
-\frac{1}{2} \, M_{H_{1}}^{2} H_{1}^{2}
+\frac{1}{2} \, m_{B^{\pm}}^{2} \, B_{i}^{+} \, B_{i}^{\, -}
\nonumber \\
&& +\frac{1}{2} \, m_{G_{\mu}}^{2} G_{\mu} \, G^{\mu}
\!+\frac{u^{2}}{4} \left[ \, g_{2} \, \cos\beta \, G\,'_{i}
+ \frac{1}{2} \left( \, g^{\prime} \, Y\,'_{i} - g_{2} \, B_{i}^{\, 3} \, \right) \right]^{2} \; ,
\end{eqnarray}
where we can obtain the mass of $G^{\mu}$ given by the expression
\begin{eqnarray}\label{massesGmuGmunu}
m_{G^{\mu}}=u \, \sqrt{g_{1}^{2} \, + \, g_{1}^{\prime \, 2}}=\frac{g_{1} \, u}{\cos\alpha} \; .
\end{eqnarray}
The last term in (\ref{LHiggs1Masses}) motivates us to make the second orthogonal transformation
\begin{eqnarray}\label{transfBYmunuZmunu}
B_{i}^{3}&=& \cos\theta_{2} \, Z\,'_{i}+ \sin\theta_{2} \, A\,'_{i}
\nonumber \\
Y\,'_{i}&=&-\sin\theta_{2} \, Z\,'_{i}+\cos\theta_{2} \, A\,'_{i} \; ,
\end{eqnarray}
where $\theta_{2}$ is another mixing angle. Imposing the condition $\tan\theta_{2}=2 \, \sin\beta$,
we obtain that
\begin{eqnarray}\label{LHiggs1MassesZG}
{\cal L}_{Scalar-0}^{(1)}\!\!\!&&=\frac{1}{2} \, \left(\partial_{\mu} \, H_{1}\right)^{2}
+\frac{\lambda^{2}}{2} \, \left(\partial_{\theta i} \, H_{1}\right)^{2}
-\frac{1}{2} \, M_{H_{1}}^{2} \, H_{1}^{2}
+\frac{1}{2} \, m_{B^{\pm}}^{2} \, B_{i}^{+} \, B^{i -}
\nonumber \\
&& +\frac{1}{2} \, m_{G_{\mu}}^{2} \, G_{\mu} \, G^{\mu}
\!+\frac{g_{2}^{2} \, u^{2}}{2} \left( \, \cos\beta \, G\,'_{i}
- \frac{1}{2} \, \sec\theta_{2} \, Z\,'_{i} \right)^{2}  \, \; .
\end{eqnarray}
We write the last term in the matrix form, the diagonalization gives us the mass of $Z\,'_{i}$
\begin{eqnarray}\label{massesZmunu}
m_{Z\,'_{i}}=\frac{\sqrt{5}}{2} \, g_{2} \, u \; ,
\end{eqnarray}
while $G\,'_{i}$ remains massless in this SSB scale. Comparing the masses of $B^{\pm}$ and $Z\,'_{i}$,
we have the relation $m_{Z\,'_{i}}=\sqrt{5} \, m_{B^{\pm}}$.
Hence, we can find all the free terms of the gauge fields which are in the model after the first SSB
\begin{eqnarray}\label{FreeLGY}
{\cal L}_{Gauge-0}&=&-\frac{1}{4} \, \left( \, \partial_{\mu} \, Y_{\nu}-\partial_{\nu} \, Y_{\mu} \, \right)^{2}
-\frac{1}{4}\left( \, \partial_{\mu} \, A_{\nu}^{3}-\partial_{\nu} \, A_{\mu}^{3} \, \right)^{2}
-\frac{1}{2} \, W_{\mu\nu}^{+} \, W^{\mu\nu \, -}
\nonumber \\
&&-\frac{1}{4} \, \left( \, \partial_{\mu} \, G_{\nu}-\partial_{\nu} \, G_{\mu} \, \right)^{2}+\frac{1}{2}\, m_{G_{\mu}}^{\, 2} G_{\mu} \, G^{\mu}
\nonumber \\
&&-\lambda^{2} \, G_{ij}^{ \, +} \, G_{ij}^{\, -}
+ m_{B^{\pm}}^{\, 2} \, B_{i}^{ \, +} \, B_{i}^{ \, -}
\nonumber \\
&&-\frac{\lambda^{2}}{2} \, \left( \, \partial_{\theta i} \, Z\,'_{j}-\partial_{\theta j} \, Z\,'_{i} \, \right)^{2}
+\frac{1}{2} \, m_{Z\,'_{i}}^{\, 2} \, Z\,'_{i} \, Z\,'_{i}
\nonumber \\
&&-\frac{\lambda^{2}}{2} \, \left( \, \partial_{\theta i} \, A\,'_{j}-\partial_{\theta j} \, A\,'_{i} \, \right)^{2}
-\frac{\lambda^{2}}{2} \, \left( \, \partial_{\theta i} \, G\,'_{j}-\partial_{\theta j} \, G\,'_{i} \, \right)^{2}
\nonumber \\
&&+ \, \mbox{mixing terms between gauge boson fields} \; ,
\end{eqnarray}
where $W_{\mu\nu}^{\pm}:=\partial_{\mu}W_{\nu}^{\pm}-\partial_{\nu}W_{\mu}^{\pm}$, and $G_{ij}^{\pm}:=\partial_{\theta i} \, B_{j}^{\, \pm}-\partial_{\theta j}\, B_{i}^{\, \pm}$.
We have the massive fields $G^{\mu}$, $B^{\pm}$ and $Z_{i}$ that are associated with the NC context of the model,
while the other fields $Y^{\mu}$, $A^{\mu\,3}$, $W^{\pm}$, $G\,'_{i}$ and $A\,'_{i}$ remained massless so far.

Using the representation (\ref{PhiH1}) of $\Phi_{1}$ in the Yukawa Lagrangian (\ref{LYukawa1}), the quadratic terms
in the fermion fields are given by
\begin{equation}\label{LYukawa0f}
{\cal L}_{Yukawa-0}^{\, (1)}= \frac{i}{2} \, \frac{u}{\sqrt{2}} \, \bar{\ell} \, \left(f_{1}\,P_{R}+f_{2} \,P_{L} \right) \lambda \overrightarrow{\Sigma} \cdot \overrightarrow{\nabla}_{\theta} \, \ell
-\frac{i}{2} \, \frac{u}{\sqrt{2}} \, \lambda \, \overrightarrow{\nabla}_{\theta} \, \bar{\ell} \, \left(f_{1}^{\ast}\,P_{R}+f_{2}^{\ast} \,P_{L} \right) \cdot \overrightarrow{\Sigma} \, \ell \; ,
\end{equation}
and imposing the condition $f_{1}=f_{1}^{\ast}=f_{2}=f_{2}^{\ast}=\sqrt{2}\,u^{-1}$, after an integration by parts, we have obtained the kinetic term
for Leptons in the $\theta$-space
\begin{equation}\label{LYukawa0}
{\cal L}_{Yukawa-0}^{\, (1)}= i \, \bar{\ell} \, \lambda \, \overrightarrow{\Sigma} \cdot \overrightarrow{\nabla}_{\theta} \, \ell \; .
\end{equation}
The interactions between leptons-neutrinos and the gauge vector bosons, as highlighted in (\ref{LintAY}), can be written in terms of the
fields $G^{\mu}$ and $Y^{\mu}$ to show us the emergence of hypercharge generators of the left-right sectors as
\begin{eqnarray}\label{LintGY}
{\cal L}_{Leptons-GY}^{\, int}\!\!\!
&&=-\bar{\Psi}_{L} \star \gamma^{\mu}\left[ \, g_{1} \, A_{\mu}^{3} \, I^{3}
+\left( \, g_{1} \, \sin\alpha-J_{L} \, g_{1}^{\prime} \, \cos\alpha \, \right) Y_{\mu} \, \right] \star \Psi_{L}
\nonumber \\
&&
+\bar{\ell}_{R}\star\gamma^{\mu}\left( \, J_{R} \, g_{1}^{\prime} \, \cos\alpha \, Y_{\mu} \, \right)\star \ell_{R}
+\bar{\Psi}_{L}\star \gamma^{\mu}\left[ \!\!\! \phantom{\frac{1}{2}} \Psi_{L} \, , \, J_{L} \, g_{1}^{\prime} \, \cos\alpha \, Y_{\mu} \, \right]_{\star}
\nonumber \\
&&
+\bar{\ell}_{R}\star \gamma^{\mu}\left[ \!\!\! \phantom{\frac{1}{2}} \ell_{R} \, , \, J_{R} \, g_{1}^{\prime} \, \cos\alpha \, Y_{\mu} \, \right]_{\star}
-\bar{\Psi}_{L} \star \gamma^{\mu}\left( \, g_{1} \, \cos\alpha + J_{L} \, g_{1}^{\prime} \, \sin\alpha \, \right) \, G_{\mu} \star \Psi_{L}
\nonumber \\
&&
+\bar{\ell}_{R}\star\gamma^{\mu}\left( \, -J_{R} \, g_{1}^{\prime} \, \sin\alpha \, G_{\mu} \, \right)\star \ell_{R}
+\bar{\Psi}_{L}\star \gamma^{\mu}\left[ \!\!\! \phantom{\frac{1}{2}} \Psi_{L} \, , \, -J_{L} \, g_{1}^{\prime} \, \sin\alpha \, G_{\mu} \right]_{\star}
\nonumber \\
&&
+\bar{\ell}_{R}\star \gamma^{\mu}\left[ \!\!\! \phantom{\frac{1}{2}} \ell_{R} \, , \, -J_{R} \, g_{1}^{\prime} \, \sin\alpha \, G_{\mu}  \right]_{\star}
\; .
\end{eqnarray}
This expression suggests us to define the hypercharge right and left-handed as being
\begin{eqnarray}
g \, Y_{R} = - \, J_{R} \, g_{1}^{\prime} \, \cos\alpha
\hspace{0.5cm} \mbox{and} \hspace{0.5cm}
g \, Y_{L} = g_{1} \, \sin\alpha - J_{L} \, g_{1}^{\prime} \, \cos\alpha \; ,
\end{eqnarray}
and $g$ is the coupling constant, attached to the hypercharge generator $Y$ of left or right sectors,
after this first SSB. Using the relation (\ref{gg1g1prime}), we obtain
\begin{eqnarray}\label{YJ}
Y_{R} = - \, J_{R}
\hspace{0.5cm} \mbox{and} \hspace{0.5cm}
Y_{L} = 1 - J_{L} \; .
\end{eqnarray}
These definitions give us
\begin{eqnarray}\label{LintGY}
{\cal L}_{Leptons-GY}^{\, int}\!\!\!\!&&=-\bar{\Psi}_{L} \star \left( \, g_{1} \, \slash{\!\!\!\!A}^{3} \, I^{3}
+ g \, Y_{L} \, {\slash \!\!\!Y} \, \right) \star \Psi_{L}
+\bar{\ell}_{R}\star \left( \, - g \, Y_{R} \, \slash{\!\!\! Y} \, \right) \star \ell_{R}
\nonumber \\
&& +\bar{\Psi}_{L} \star \gamma^{\mu}\left[ \!\!\! \phantom{\frac{1}{2}} \Psi_{L} \, , \, g \, \left(1- \, Y_{L} \, \right) \, Y_{\mu} \right]_{\star}
\!+\bar{\ell}_{R} \star \gamma^{\mu}\left[ \!\!\! \phantom{\frac{1}{2}} \ell_{R} \, , \, \left(-g \, Y_{R} \right) Y_{\mu} \, \right]_{\star}
\nonumber \\
&&
-\bar{\Psi}_{L} \star \frac{g_{1}}{\cos\alpha} \left( \, 1- Y_{L} \, \sin^2\alpha \, \right) \, \slash{\!\!\!\!G} \star \Psi_{L}
+\bar{\ell}_{R}\star\left( \, g \, Y_{R} \, \tan\alpha \,\, \slash{\!\!\!\!G} \, \right)\star \ell_{R}
\nonumber \\
&&
+\bar{\Psi}_{L}\star \gamma^{\mu}\left[ \!\!\! \phantom{\frac{1}{2}} \Psi_{L} \, , \, g \, \tan\alpha
\left(- \, 1 + Y_{L} \, \right) G_{\mu} \right]_{\star}
\nonumber \\
&&
+\bar{\ell}_{R}\star \gamma^{\mu}\left[ \!\!\! \phantom{\frac{1}{2}} \ell_{R} \, , \, g \, Y_{R} \, \tan\alpha \, G_{\mu} \, \right]_{\star} .
\end{eqnarray}
Here we are ready to discover how the mixing $A^{\mu \, 3}-Y^{\mu}$ defines the physical particles $Z^{0}$ and the massless photon,
and after that, the electric charge of the particles. To accomplish this task, we need to break the residual electroweak symmetry
\begin{eqnarray}
SU_{L}^{\star}(2)_{A^{\mu a}} \times U_{Y}^{\star}(1) \times U^{\star}(1)_{G\,'_{i}} \times U_{R}^{\star}(1)_{Y'_{i}} \; .
\end{eqnarray}
We will carry out this second mechanism in the next section.

\section{The electroweak symmetry breaking}
\renewcommand{\theequation}{5.\arabic{equation}}
\setcounter{equation}{0}

In the previous section we have constructed a model for the symmetry $U_{L}^{\star}(2)_{A^{\mu}} \times U_{R}^{\star}(1)_{X^{\mu}}
\times U_{L}^{\star}(2)_{B_{i}} \times U_{R}^{\star}(1)_{C_{i}}$, and we have used a Higgs sector to eliminate the residual symmetry $U^{\star}(1)_{A^{\mu0}}$,
and we have defined the hypercharge of the Abelian sector, remaining the NC electroweak symmetry
$SU_{L}^{\star}(2)_{A^{\mu \, a}} \times U_{Y}^{\star}(1) \times U^{\star}(1)_{G\,'_{i}} \times U_{R}^{\star}(1)_{C_{i}}$. Now we are going to introduce a second Higgs sector $\Phi_{2}$ in order to break the
electroweak symmetry. We write the Lagrangian of the second Higgs-$\Phi_{2}$ as the sum of the scalar sector and the Yukawa Lagrangian, that is,
\begin{eqnarray}\label{LHiggs2}
{\cal L}_{Higgs}^{\, (2)}={\cal L}_{Scalar}^{\, (2)}+{\cal L}_{Yukawa}^{ \, (2)} \; ,
\end{eqnarray}
where we have that
\begin{equation}\label{LScalar2}
{\cal L}_{Scalar}^{\,(2)}=\left(D_{\mu}\Phi_{2}\right)^{\dagger} \star D^{\mu} \Phi_{2}
+ \left(D_{\theta i} \Phi_{2} \right)^{\dagger} \star D_{\theta i}\Phi_{2}
-\mu_{2}^2 \, \left(\Phi_{2}^{\dagger} \star \Phi_{2}\right)-g_{H2} \, \left(\Phi_{2}^{\dagger} \star \Phi_{2}\right)^{2} \; ,
\end{equation}
and
\begin{eqnarray}\label{LYukawa2}
{\cal L}_{Yukawa}^{\,(2)}=- \, f_{\ell} \, \bar{\Psi}_{L} \star \Phi_{2} \star \ell_{R}
- \, f_{\ell}^{\ast} \, \bar{\ell}_{R} \star \Phi_{2}^{\dagger} \star \Psi_{L} \; .
\end{eqnarray}
Here, $\mu_{2}$ and $g_{H2}$ are real parameters, and $f_{\ell} \, (\ell=e,\mu,\tau)$ are complex ones.
The field $\Phi_{2}$ is a complex scalar doublet that has the gauge transformation under $U_{L}^{\star}(2)$ given by
\begin{eqnarray}\label{transfPhi2}
\Phi_{2}=
\left(
\begin{array}{c}
\phi_{2}^{ \, (+)} \\
\phi_{2}^{\, (0)} \\
\end{array}
\right)
\longmapsto \Phi_{2}^{\prime}=U \star \Phi_{2} \; ,
\end{eqnarray}
where $U$ is the element of $U_{L}^{\star}(2)$. The covariant derivatives act on $\Phi_{2}$  according to
\begin{eqnarray}\label{DmuPhi2}
D_{\mu} \, \Phi_{2} \!&=&\! \partial_{\mu}\Phi_{2} + i \, g_{1} \, A_{\mu}^{0} \star \Phi_{2}
+ i \, g_{1} \, A_{\mu}^{\, a} \, \frac{\sigma^{a}}{2} \star \Phi_{2} \; ,
\nonumber \\
D_{\theta i} \, \Phi_{2} \!&=&\! \lambda \, \partial_{\theta i}\Phi_{2} + i \, g_{2} \, B_{i}^{\, 0} \star \Phi_{2}
-i \, J_{\Phi_{2}} \, g_{2}^{\prime} \, \Phi_{2} \star C_{i} \; .
\end{eqnarray}
Using the transformation (\ref{transfB0X}) into the covariant derivative (\ref{DmuPhi2}), the hypercharge of $\Phi_{2}$ is
identified as
\begin{eqnarray}\label{YLRBosonsAntiSim}
g^{\prime} \, Y_{\Phi_{2}}=g_{2} \, \sin\beta - J_{\Phi_{2}} \, g_{2}^{\prime} \, \cos\beta \; ,
\end{eqnarray}
where the coupling constant $g^{\prime}$ is attached to the hypercharge generator.

%

%

%
As before, we have used a second VEV, i.e., $v \neq u$, where $u \gg v $ and which breaks the electroweak symmetry
of (\ref{LHiggs2}).
%
%
We can choose the parametrization of the $\Phi_{2}$-complex field as being
\begin{eqnarray}\label{PhiGaugeunitary}
\Phi_{2}(x,\theta)= \frac{\left(v+H_{2}\right)}{\sqrt{2}} \star  e^{i\frac{\sigma^{a}}{2}\frac{\chi^{a}}{v}}
\left(
  \begin{array}{c}
    0 \\
    1 \\
  \end{array}
\right) \; ,
\end{eqnarray}
where, in the unitary gauge, can be written as
%
\begin{eqnarray}\label{PhiH}
\Phi_{2}(x,\theta)=
\frac{v+H_{2}(x,\theta)}{\sqrt{2}}
\left(
  \begin{array}{c}
    0 \\
    1 \\
  \end{array}
\right) \; .
\end{eqnarray}
The free terms in the scalar sector of the Lagrangian (\ref{LHiggs2}) are given by
\begin{eqnarray}\label{massesBWZ}
&&{\cal L}_{Scalar-0}^{\,(2)}=\frac{1}{2}\,\left(\partial_{\mu}H_{2}\right)^{2}
+\frac{\lambda^{2}}{2}\left(\partial_{\theta \, i}H_{2}\right)^{2}
-\frac{1}{2}\left(-2\mu^{2}\right)H_{2}^{2}
+\frac{g_{1}^{2}v^{2}}{4}\, W_{\mu}^{\;+}W^{\mu-}
\nonumber \\
&&+\frac{v^{2}}{2} \left[ \, g_{1} \, \cos\alpha \, G_{\mu}+\frac{1}{2} \left( \phantom{\frac{1}{2}} \!\!\!\! g \, Y_{\mu}-g_{1} \, A_{\mu}^{3} \right) \right]^{2}
\!+\frac{v^{2}}{4} \left( \!\!\!\!\! \phantom{\frac{1}{2}} \, g_{2} \, B_{i}^{\, 0}- J_{\Phi_{2}} \, g_{2}^{\prime} \, C_{i} \right)^{2} ,
\hspace{1.2cm}
\end{eqnarray}
where we have used the Moyal product (\ref{ProductMoyal}). In this expression we can obtain
the Higgs scalar $H_{2}$-field Lagrangian with propagation in the $\theta$-space, and a mass of $M_{H_{2}}=\sqrt{2 \, g_{2H} \, v^{\, 2}}$.
As in the usual case, the mass of $W^{\pm}$ is
\begin{eqnarray}
m_{W^{\pm}}= \frac{1}{2} \, g_{1} \, v \; .
\end{eqnarray}
%
%
%
The mass terms of the neutral bosons in (\ref{massesBWZ}) motivate us to introduce the orthogonal transformations
\begin{eqnarray}\label{transfAYZ}
A_{\mu}^{3}&=& \cos\theta_{W} \, Z_{\mu} + \sin\theta_{W} \, A_{\mu}
\nonumber \\
Y_{\mu}&=&-\sin\theta_{W} \, Z_{\mu} + \cos\theta_{W} \, A_{\mu} \; ,
\end{eqnarray}
%
%
where $\theta_{W}$ is the Weinberg's angle, and using the relation (\ref{tanalpha}), we impose that
\begin{eqnarray}\label{tanalphabeta}
\tan\theta_{W}=\frac{g}{g_{1}}
\hspace{0.5cm} \mbox{and} \hspace{0.5cm}
\tan\beta=\frac{g_{2}^{\prime}}{g_{2}} \; .
\end{eqnarray}
Thus the definition (\ref{YLRBosonsAntiSim}) confirms the relation $Y_{\Phi_{2}}=1-J_{\Phi_{2}}$, and the coupling constant $g^{\prime}$
is parameterized by
\begin{eqnarray}
g^{\, \prime}=g_{2} \, \sin\beta=g_{2}^{\, \prime} \, \cos\beta \; .
\end{eqnarray}
We meet all the massive terms of the model
\begin{eqnarray}\label{Lmass}
{\cal L}_{mass}&=&m_{W^{\pm}}^{2}\, W_{\mu}^{\;+}W^{\mu-}
\!+\frac{g_{1}^{2} \, v^{2}}{2} \left( \, \frac{1}{2} \, \sec\theta_{W} \, Z_{\mu} - \cos\alpha \, G_{\mu} \, \right)^{2}
\!+\frac{1}{2} \, \frac{g_{1}^{2} \, u^{2}}{\cos^{2}\alpha}  \, G_{\mu}G^{\mu}
\nonumber \\
&&+ m_{B^{\pm}}^{2} \, B_{i}^{\, +} \, B_{i}^{\, -}\!
+\frac{1}{2} \, \frac{g_{2}^{2} \, v^{2}}{\cos^2\beta} \, G\,'_{i} \, G\,'_{i}\!+\frac{1}{2} \, \frac{5 \, g_{2}^{2} \, u^{2}}{4}  \, Z\,'_{i} \, Z\,'_{i} \; .
\end{eqnarray}
Here we have taken into account the mass terms from the first SSB of the Higgs-$\Phi_{1}$.
We can observe that the fields $A^{\mu}$ and $A\,'_{i}$ are not present in the Lagrangian (\ref{massesBWZ}).
They are the massless gauge fields remaining in the model after SSBs, that is, we have the final symmetry
\begin{eqnarray}
U^{\star}(1)_{A^{\mu0}} \times SU_{L}^{\star}(2)_{A^{\mu \, a}} \times U_{R}^{\star}(1)_{X^{\mu}} \times U^{\star}(1)_{B_{i}^{ \, 0}} \times
SU_{L}^{\star}(2)_{B_{i}^{\, a}}  \times U_{R}^{\star}(1)_{C_{i}}
\nonumber \\
\stackrel{\langle \Phi_{1} \rangle}{\longmapsto} SU_{L}^{\star}(2) \times U_{Y}^{\star}(1) \times U^{\star}(1)_{G\,'_{i}} \times U_{R}^{\star}(1)_{C_{i}}
\nonumber \\
\stackrel{\langle \Phi_{2}
\rangle}{\longmapsto} U_{em}^{\star}(1)\times U^{\star}(1)_{A\,'_{i}} \; ,
\end{eqnarray}
where $A^{\mu}$ is the NC photon field and
$A\,'_{i}$ is its antisymmetrical correspondent in the $\theta$-space.
It is important to explain that the $Z$-field, which came from (\ref{transfAYZ}) is not the $Z^{0}$-particle of the standard electroweak model.
The $Z^{0}$-particle will be defined by means of the mixing with the $G$-field in (\ref{Lmass}). To complete the task, we will write the $Z^{\mu}-G^{\mu}$ massive term of (\ref{Lmass}) in a matrix form
\begin{eqnarray}\label{LmassGZ}
{\cal L}_{mass \, (Z^{\mu}-G^{\mu})}&&\!\!=\frac{g_{1}^{2}v^{2}}{8} \sec^{2}\theta_{W}
\left(
  \begin{array}{cc}
    Z_{\mu} & G_{\mu} \\
  \end{array}
\right)
\left(
\begin{array}{cc}
1 & -a \\
-a & a^2+b^{2} \\
  \end{array}
\right)
\left(
\begin{array}{c}
Z^{\mu} \\
G^{\mu} \\
\end{array}
\right) \; ,
\end{eqnarray}
where, for simplicity, we have defined the real constants $a=2\cos\theta_{W} \cos\alpha$, and
$b=2 \, \frac{u}{v} \, \cos\theta_{W} \sec\alpha$. Since we had established the scale $u \gg v$,
we have diagonalized the previous matrices of (\ref{LmassGZ}) under this
condition, so the masses of $Z$, $G$ and their antisymmetric pairs up to second order in $v/u$ can be given by
\begin{eqnarray}
m_{Z^{0}}\! &=& \!\frac{g_{1}v}{2\cos\theta_{W}} \left(1-\frac{v^{2}}{2 u^{2}} \cos^{4}\alpha + \ldots  \right)
\; ,
\nonumber \\
m_{G^{\mu}}\!&=&\!\frac{g_{1}u}{\cos\alpha}\left(1+\frac{v^{2}}{2 u^{2}} \cos^{4}\alpha +\ldots  \right) \; ,
\nonumber \\
m_{G'_{i}}\!&=&\!\frac{g_{2}v}{\cos\beta\sec\theta_{2}}\left(1-\frac{2v^{2}}{u^{2}} \sec^{4}\beta \cos^{4}\theta_{2} + \ldots \right) \; .
\end{eqnarray}

Now we will see how the electromagnetic and weak interactions can appear in the Lagrangian (\ref{LintGY}).
Substituting (\ref{transfAYZ}) into (\ref{LintGY}), we can identify the fundamental charge by the parametrization
\begin{eqnarray}\label{eg1g}
e=g_{1}\sin\theta_{W}=g\cos\theta_{W} \; ,
\end{eqnarray}
where the electric charge is given by
\begin{eqnarray}\label{Nishijima}
Q_{em}=I^{3}+Y \; .
\end{eqnarray}
We summarize the values of $I^{3}$, $Y$ and $Q_{em}$ in the following table
\begin{eqnarray*}
\begin{tabular}{|l|l|l|l|l|l|}
\hline
\mbox{Fields} \& \mbox{Particles} & $Q_{em}\!=\!I^{3}+Y$ & $I^{3}$ & $Y\!=\!1-J$ & $Y=-J$ & $J$ \\
\hline
\mbox{Leptons-left} & $-1$ & $-1/2$ & $-1/2$ & Not apply & $+3/2$  \\
\hline
\mbox{Leptons-right} & $-1$ & $0$ & Not apply & $-1$ & $+1$ \\
\hline
\mbox{Neutrinos-left} & $0$ & $+1/2$ & $-1/2$ & Not apply & $ +3/2 $ \\
\hline
\mbox{Neutrino-right} & $0$ & $0$ & Not apply & 0 & $0$ \\
\hline
\mbox{Bosons} $W^{\pm}$ & $\pm \, 1$ & $\pm \, 1$ & $0$ & Not apply & $+1$ \\
\hline
\mbox{Neutral Bosons} & $0$ & $0$ & $0$ & Not apply & $+1$ \\
\hline
\mbox{Higgs Singlet} & $0$ & $0$ & $0$ & Not apply & $+1$ \\
\hline
\mbox{Higgs Doublet} & $0$ & $-1/2$ & $+1/2$ & Not apply & $+1/2$ \\
\hline
\end{tabular}
\end{eqnarray*}

\vspace{0.3cm}

Using the previous results, we can obtain a relation between the masses of the charged bosons $W^{\pm}$ and the neutral bosons $Z^{\mu}-G^{\mu}$
due to the contribution of the NC space-time
\begin{eqnarray}\label{ratiomm}
\frac{m^{2}_{W^{\pm}}}{m^{2}_{Z^{0}}}= \cos^{2}\theta_{W} \left[1+ 4 \left(\frac{m_{Z^{0}}}{m_{G^{\mu}}}\right)^{2}
\cos^2\theta_{W} \cos^{2}\alpha + \, \ldots  \right]
 \; .
\end{eqnarray}
The parametrization (\ref{eg1g}) gives us the masses of $W^{\pm}$, $Z^{0}$ and $G^{\mu}$ in terms of $e$, $v$
and the mixing angle $\theta_{W}$
\begin{eqnarray}
m_{W^{\pm}}&=&\frac{e \, v }{2\sin\theta_{W}} \; ,
\nonumber \\
m_{Z^{0}}&=&\frac{e\, v}{\sin\left(2\theta_{W} \right)} \left(1-\frac{v^{2}}{2 u^{2}} \cos^{4}\alpha + \ldots  \right) \; ,
\nonumber \\
m_{G^{\mu}}&=&\frac{e\, u}{\sin\theta_{W}\cos\alpha} \left(1+\frac{v^{2}}{2 u^{2}} \cos^{4}\alpha +\ldots  \right) \; .
\end{eqnarray}
The relation between $g_{1}$ and the Fermi's constant $G_{F}=1.166 \, \times \, 10^{-5} \left(\mbox{GeV}\right)^{-2}$, i.e., $g_{1}^{2}=4\sqrt{2}\, m_{W^{\pm}}^{\,2} \, G_{F}$, leads us to $v=\left(\sqrt{2} \, G_{F}\right)^{-1/2}\simeq 246 \, \mbox{GeV}$,
and hence the masses of $W^{\pm}$ and $Z^{0}$ are given by
\begin{eqnarray}\label{massesWZ}
m_{W^{\pm}}=\frac{37 \, \, \mbox{GeV}}{|\sin\theta_{W}|}
\hspace{0.5cm} , \hspace{0.5cm}
m_{Z^{0}}=\frac{74 \, \, \mbox{GeV}}{|\sin2\theta_{W}|} \left(1-\frac{v^{2}}{2 u^{2}} \cos^{4}\alpha + \ldots  \right)  \; .
\end{eqnarray}
Since we identify the $\theta_{W}$-angle as the Weinberg angle, considering that the experimental value of
$\sin^{2}\theta_{W}$ is around $\simeq 0.23$, the masses of $W^{\pm}$ and $Z^{0}$
are estimated to give the values
\begin{eqnarray}\label{mWmZ}
m_{W^{\pm}}\simeq 77 \, \mbox{GeV}
\hspace{0.5cm} \mbox{and} \hspace{0.5cm}
m_{Z^{0}}\simeq 89 \, \mbox{GeV} \, \left(1-\frac{v^{2}}{2 u^{2}} \cos^{4}\alpha + \ldots  \right) \; ,
\end{eqnarray}
in which the $\alpha$-angle is a contribution of the NC space-time.

Finally, we can find all the massive terms of the free Lagrangian (\ref{FreeLGY}) as
\begin{eqnarray}\label{FreeLGY}
{\cal L}_{Gauge-0}&=&-\frac{1}{4} \, \left(\partial_{\mu}A_{\nu}-\partial_{\nu}A_{\mu}\right)^{2}
-\frac{1}{2} \, W_{\mu\nu}^{+}W^{\mu\nu \, -}+m_{W^{\pm}}^{\, 2} W_{\mu}^{\,+}W^{\mu \, -}
\nonumber \\
&&-\frac{1}{4}\left(\partial_{\mu}Z_{\nu}-\partial_{\nu}Z_{\mu} \right)^{2}+\frac{1}{2}\, m_{Z^{0}}^{\, 2} Z_{\mu}\,Z^{\mu}
\nonumber \\
&&-\frac{1}{4} \, \left(\partial_{\mu}G_{\nu}-\partial_{\nu}G_{\mu} \right)^{2}+\frac{1}{2}\, m_{G_{\mu}}^{\, 2} G_{\mu}\,G^{\mu}
\nonumber \\
&&-\frac{\lambda^{2}}{2} \, \left(\partial_{\theta i}A'_{j}-\partial_{\theta j}A'_{i}\right)^{2}
-\frac{\lambda^{2}}{2} \, G_{ij}^{\, +}G_{ij}^{\, -}
+ m_{B^{\pm}}^{\, 2} \, B_{i}^{\, +} \, B_{i}^{\, -}
\nonumber \\
&&-\frac{\lambda^{2}}{2} \, \left(\partial_{\theta i}Z'_{j}-\partial_{\theta j}Z'_{i} \right)^{2}
+\frac{1}{2} \, m_{Z'_{i}}^{\, 2} \, Z'_{i}\, Z'_{i}
\nonumber \\
&&
-\frac{\lambda^{2}}{2} \, \left(\partial_{\theta i} \, G'_{j}-\partial_{\theta j} \, G'_{i}\right)^{2}
+\frac{1}{2} \, m_{G_{i}}^{\, 2} \, G'_{i} \, G'_{i}
\nonumber \\
&&+ \, \mbox{mixing terms between gauge boson fields} \; ,
\end{eqnarray}
where only the gauge fields $A^{\mu}$ and $A'_{i}$ are massless in this last expression.

In order to obtain a numerical value for the masses of $G^{\mu} , B^{\pm}_{i} , Z'_{i}$ and $G'_{i}$ in terms of the electric charge
and the mixing angles, we need to analyze the self-interactions of the gauge bosons.  This issue will be discussed in the next section.

\section{The interaction sectors and numerical masses of the new bosons}
\renewcommand{\theequation}{6.\arabic{equation}}
\setcounter{equation}{0}

The Higgs sector introduced in the previous section has shown us the physical fields, such that the massless photon
and other massive bosons of the model. We identify the electromagnetic interaction through the Nishijima relation (\ref{Nishijima})
and the parametrization (\ref{eg1g}). Using these expressions, the interactions of gauge bosons and leptons-neutrinos in (\ref{LintGY})
can be written as
\begin{eqnarray}
&&{\cal L}_{Leptons-Gauge}^{\, int}
=-\frac{g_{1}}{2\sqrt{2}} \, \bar{\nu_{\ell}} \star \left(1+\gamma_{5}\right) \, \slash{\!\!\!\!W}^{+} \star\ell
-\frac{g_{1}}{2\sqrt{2}} \, \bar{\ell} \star \left(1+\gamma_{5}\right) \, \slash{\!\!\!\!W}^{-}  \star \nu_{\ell}
\nonumber \\
&&-e \, Q_{em} \, \bar{\Psi}_{L}\star \, \slash{\!\!\!\! A} \star \Psi_{L}
- e \, Y_{R} \, \bar{\ell}_{R}\star \, \slash{\!\!\!\! A} \star \ell_{R}
\nonumber \\
&&+\left( \, 1- Y_{L} \, \right)g \, \cos\theta_{W} \, \bar{\Psi}_{L}\star \gamma^{\mu}\left[ \!\!\! \phantom{\frac{1}{2}}  \Psi_{L} \, , \, A_{\mu} \, \right]_{\star}
\!\!-e \, Y_{R} \, \bar{\ell}_{R} \star \gamma^{\mu}\left[ \!\!\! \phantom{\frac{1}{2}} \ell_{R} \, , \, A_{\mu} \, \right]_{\star}
\nonumber \\
&&-\bar{\Psi}_{L}\star g_{1} \, \cos\theta_{W} \left( \, I^{3}- \tan^{2}\theta_{W} \, Y_{L} \, \right) \, \slash{\!\!\!\! Z} \star \Psi_{L}
+g \, Y_{R} \, \sin\theta_{W} \, \bar{\ell}_{R}\star \, \slash{\!\!\!\! Z} \star \ell_{R}
\nonumber \\
&&-g \, \left( \, 1 - \, Y_{L} \,\right) \, \sin\theta_{W} \, \bar{\Psi}_{L}\star \gamma^{\mu}\left[ \!\!\! \phantom{\frac{1}{2}} \Psi_{L} \,, \,Z_{\mu} \, \right]_{\star}
\!\!+g \, Y_{R} \, \sin\theta_{W} \, \bar{\ell}_{R}\star \gamma^{\mu}\left[ \!\!\! \phantom{\frac{1}{2}} \ell_{R} \, , \, Z_{\mu} \, \right]_{\star}
\nonumber \\
&&-\bar{\Psi}_{L} \star \frac{g_{1}}{\cos\alpha} \left( \, 1- Y_{L} \, \sin^2\alpha \, \right) \, \slash{\!\!\!\! G} \star \Psi_{L}
+\bar{\ell}_{R}\star\left( \, g \, Y_{R} \, \tan\alpha \, \, \slash{\!\!\!\! G} \, \right)\star \ell_{R}
\nonumber \\
&&
+\bar{\Psi}_{L}\star \gamma^{\mu} \left[ \!\!\! \phantom{\frac{1}{2}} \Psi_{L} \, , \, g \, \tan\alpha
\left( \, -1 + Y_{L} \, \right) G_{\mu} \, \right]_{\star}
\!\!+\bar{\ell}_{R}\star \gamma^{\mu}\left[ \!\!\! \phantom{\frac{1}{2}} \ell_{R} \, , \, g \, Y_{R} \, \tan\alpha \, G_{\mu}  \right]_{\star}
 . \hspace{1.2cm}
\end{eqnarray}

To obtain the expressions for the masses of $Z'_{i}$, $B^{\pm}$ and $G'_{i}$ in terms of the fundamental parameters, firstly we have to examine the $3$-line and $4$-line vertex of the bosons $B^{\pm}$ interacting with the electromagnetic photon-$A^{\mu}$ :
%
%
%
%
\begin{figure}[!h]
\begin{center}
\newpsobject{showgrid}{psgrid}{subgriddiv=1,griddots=10,gridlabels=6pt}
\begin{pspicture}(-1,1)(14,3.1)
\psset{arrowsize=0.2 2}
\psset{unit=1}
%
%
%
\pscoil[coilarm=0,coilaspect=0,coilwidth=0.2,coilheight=1.0,linecolor=black](-1,1)(2,1)
\pscoil[coilarm=0,coilaspect=0,coilwidth=0.2,coilheight=1.0,linecolor=black](0.45,1.1)(0.45,3.1)
\put(0.7,2.8){\large$\gamma$}
\put(-1,1.2){\large$B^{+}$}
\put(1.7,1.2){\large$B^{-}$}
%
%
%
\pscoil[coilarm=0,coilaspect=0,coilwidth=0.2,coilheight=1.0,linecolor=black](3,1)(6,1)
\pscoil[coilarm=0,coilaspect=0,coilwidth=0.15,coilheight=1.0,linecolor=black](4.45,1.1)(3.5,3)
\pscoil[coilarm=0,coilaspect=0,coilwidth=0.15,coilheight=1.0,linecolor=black](4.45,1.1)(5.5,3)
\put(3.1,2.8){\large$\gamma$}
\put(5.7,2.8){\large$\gamma$}
\put(2.9,1.2){\large$B^{+}$}
\put(5.8,1.2){\large$B^{-}$}
%
%
\pscoil[coilarm=0,coilaspect=0,coilwidth=0.2,coilheight=1.0,linecolor=black](7,1)(10,1)
\pscoil[coilarm=0,coilaspect=0,coilwidth=0.2,coilheight=1.0,linecolor=black](8.45,1.1)(8.45,3.1)
\put(8.7,2.8){\large$\gamma$}
\put(6.9,1.2){\large$B^{+}$}
\put(9.7,1.2){\large$W^{-}$}
%
%
%
\pscoil[coilarm=0,coilaspect=0,coilwidth=0.2,coilheight=1.0,linecolor=black](11,1)(14,1)
\pscoil[coilarm=0,coilaspect=0,coilwidth=0.2,coilheight=1.0,linecolor=black](12.45,1.1)(13.35,3)
\pscoil[coilarm=0,coilaspect=0,coilwidth=0.15,coilheight=1.0,linecolor=black](12.45,1.1)(11.5,3)
\put(11,2.8){\large$\gamma$}
\put(13.6,2.8){\large$\gamma$}
\put(10.9,1.2){\large$B^{-}$}
\put(13.8,1.2){\large$W^{+}$}
%
%
%
\end{pspicture}
%
%
%
\end{center}
\end{figure}

\noindent
From Lagrangian sector of bosons $B^{\pm}$, we list all interactions with photon-$A^{\mu}$ :
\begin{eqnarray}
{\cal L}_{\gamma-B^{\pm}}=- \, \frac{i \, g_{2}}{2} \, \sin\theta_{W} \, \left(\partial_{i}A_{j}-\partial_{j}A_{i}\right)
\left\{ \, B^{i \,+} \, , \, B^{j \, -} \, \right\}_{\star}
\nonumber \\
-\frac{ig_{2}}{2} \, \cos\theta_{W} \sin\alpha \left(\partial_{i}A_{j}-\partial_{j}A_{i}\right) \left[ \, B^{i \, +} \, , \, B^{i \,  -} \, \right]_{\star} \,
\; ,
\end{eqnarray}
\begin{eqnarray}
{\cal L}_{\gamma\gamma-B^{\pm}}= g_{1} \, g_{2} \sin\theta_{W} \cos\theta_{W} \sin\alpha \,
\left\{B^{i \, +}, B^{j \, -} \right\}_{\star}
\, \left[ \, A_{i} \, , \, A_{j} \, \right]_{\star}
\nonumber \\
+\frac{1}{2} \, g_{1} \, g_{2} \, \left(\cos^{2}\theta_{W}+\frac{1}{4} \, \sin^{2}\alpha \sin^{2}\theta_{W} \right)
\left[ \, A_{i} \, , \, A_{j} \, \right]_{\star} \left[ \, B^{i \, +} \, , \, B^{j \, -} \, \right]_{\star}
 \; ,
\end{eqnarray}
\begin{eqnarray}
{\cal L}_{\gamma-B^{+}W^{-}}=-\frac{ig_{1}}{2} \, \sin\theta_{W} \,
\left(\lambda \partial_{\theta \, i}B_{j}^{\, +}-\lambda \partial_{\theta \, j}B_{i}^{\, +}  \right)
\left\{ \, W^{i\, -} \, , \, A^{j} \, \right\}_{\star}
\nonumber \\
-ig_{1} \cos\theta_{W} \sin\alpha \,  \left(\lambda \partial_{\theta \, i}B_{j}^{\, +}-\lambda \partial_{\theta \, j}B_{i}^{\, +}  \right)
\left[ \, W^{i\, -} \, , \, A^{j} \, \right]_{\star} \; ,
\end{eqnarray}
\begin{eqnarray}
{\cal L}_{\gamma-B^{-}W^{+}}=\frac{ig_{1}}{2} \, \sin\theta_{W} \,
\left(\lambda \partial_{\mu\rho}B_{\nu}^{\; \; \rho \, -}-\lambda \partial_{\nu}^{\; \; \rho}B_{\mu\rho}^{\; \; -}  \right)
\left\{W^{\mu +},A^{\nu} \right\}_{\star}
\nonumber \\
-ig_{1} \cos\theta_{W} \sin\alpha \,  \left(\lambda \partial_{\theta \, i}B_{j}^{\, -}-\lambda \partial_{\theta \, j}B_{i}^{\, -}  \right)
\left[ \, W^{i \, +} \, , \, A^{j} \, \right]_{\star} \; .
\end{eqnarray}

\noindent
Using the universality of the electromagnetic interaction, all previous vertex, say $\gamma-B^{\pm}$, $\gamma \, \gamma-B^{\pm}$ and $\gamma-B^{+}W^{-}$, have the fundamental electric charge $e$ as its coupling constant, where the fine structure constant is $e^{\,2}\simeq4\pi/137$. Therefore, we find an expression that connect the coupling constant $g_{2}$ to both the fundamental electric charge $e$ and the Weinberg's angle $\theta_{W}$
\begin{eqnarray}\label{g2e}
g_{2}=\frac{e}{\sin\theta_{W}} \; ,
\end{eqnarray}
and we also obtain the $\alpha$-angle connected to $\theta_{W}$, by $\sin\alpha=\tan\theta_{W}$. Using the value of the standard model
for the masses ratio in (\ref{ratiomm}), the mixing angle-$\alpha$ is given by $\sin^{2}\alpha \simeq 0.33$, see \cite{Chaichian2003},
where the relation $\sin\alpha=\tan\theta_{W}$ will be satisfied. In the NC model, the scale of NCY has a lower bound at
$\Lambda_{NC} \gtrsim 1 \, \mbox{TeV}$.   So we use this scale to represent the first $VEV$, that is, $u \simeq 1 \, \mbox{TeV}$.
Consequently, the masses of the bosons $B^{\pm}$ and $Z_{i}'$ can be computed as
\begin{eqnarray}\label{massesGBZG'}
m_{G^{\mu}} \!&=&\! \frac{e \, u}{\sin\theta_{W}\cos\alpha} \simeq 770 \, \mbox{GeV} \; ,
\nonumber \\
m_{B^{\pm}} \!&=&\! \frac{e \, u}{2\sin\theta_{W}} \simeq 310 \, \mbox{GeV} \; ,
\nonumber \\
m_{Z'} \!&=&\! \frac{\sqrt{5}}{2} \, \frac{e \, u}{\sin\theta_{W}}
\simeq 699 \, \mbox{GeV} \; ,
\nonumber \\
m_{G'} \!&=&\! \frac{e \, v}{\sin\theta_{W}\cos\beta}
\simeq \frac{154}{\cos\beta} \, \mbox{GeV} > 154 \, \mbox{GeV} \; .
\end{eqnarray}

To place our results inside an experimental scenario with the results obtained in LHC \cite{atlas}, we will compare them
with the masses obtained for the NC gauge bosons, for instance. Firstly, we will use the VEV $u$-scale as the NC
scale $\Lambda_{NC}$ estimated in the literature $ u \simeq \Lambda_{NC} = 1 \, \mbox{TeV}$. Within this scale, let us consider
the values listed in Eq. (\ref{massesGBZG'}) just below. The NC model that we are dealing with in this
work has $(3 + 1 + 3)$-dimensions, in which the $3$ extra dimensions are compactified. The bosons $B^{\pm}$ and $Z_{i}'$ are,
specifically, due to these three extra dimensions, and we will compare their masses to the $W′$ and $Z′$ in \cite{atlas}.
If we adopt the VEV-$u$ as the TeV-scale for \cite{atlas}, {\it i.e.}, $u \simeq \sqrt{s} = 8 \, \mbox{TeV}$ ,
we can obtain the following masses for our results in (\ref{massesGBZG'})
\begin{eqnarray}\label{massesGBZG'}
m_{G^{\mu}} \!&=& \! 6.1 \, \mbox{TeV} \; ,
\nonumber \\
m_{B^{\pm}} \!&=&\! 2.5 \, \mbox{TeV} \; ,
\nonumber \\
m_{Z'} \!&=&\! 5.5 \, \mbox{TeV} \; .
\end{eqnarray}
The mass of $W^{\prime}$, and a lower limit on the mass of $Z^{\prime}$, according to ATLAS and CMS experiment, are given by
\begin{eqnarray}
m_{W^{\prime}} = 2.5 \, \mbox{TeV}
\hspace{0.5cm} \mbox{and} \hspace{0.5cm}
m_{Z^{\prime}} = 2.95 \, \mbox{TeV} \; ,
\end{eqnarray}
which means that our results in (\ref{massesGBZG'}) have the same TeV order of energy. Notice that the NC contributions increased
these numbers, which is a reasonable result.

Other important constraint is the one which analyzes the experimental uncertainty of $Z^{0}$ mass
\begin{eqnarray}\label{massZ0Exp}
m_{Z^{0}}=91.1876 \, \pm \, 0.0021 \;  \mbox{GeV} \; .
\end{eqnarray}
Using this uncertainty in the result (\ref{mWmZ}), the VEV scale-$u$ is fixed by the value $u= 2.5 \, \mbox{TeV}$.
Thereby the masses of $G^{\mu}$, $B^{\pm}$ and $Z'$ are given by
\begin{eqnarray}\label{massesGBZG'uTeV}
m_{G^{\mu}} \!&=& \! 1.93 \, \mbox{TeV} \; ,
\nonumber \\
m_{B^{\pm}} \!&=&\! 781.2 \, \mbox{GeV} \; ,
\nonumber \\
m_{Z'} \!&=&\! 1.75 \, \mbox{TeV} \; .
\end{eqnarray}

In this way, we have analyzed some elements of the NC standard model such as the electroweak standard model. Since the position and $\theta$ coordinates are independent variables, the MW product keeps its associative property.   This product is the basic usual one, in DFR phase-space in canonical NC models.

Hence, we have introduced new ideas and concepts in
DFR formalism.  Let us summarize them all.  We have begun with the construction of the
symmetry group $U_{L}^{\star}(2) \times U_{R}^{\star}(1)$, which is the DFR version
of the GSW model concerning the electroweak interaction,
in order to introduce left- and right-handed fermionic
sectors. Some elements such as covariant derivatives (QFT-kind),
gauge transformations and gauge invariant Lagrangians
were constructed.  The interactions between leptons
and gauge fields were discussed.
After that we have introduced the first Higgs sector to
break one of the two Abelian NC symmetries in order to
destroy the residual model's $U^{\star}(1)$ symmetry. The spontaneous
symmetry breaking was discussed and, in this way,
the Higgs Lagrangian was introduced. We have seen that,
in the context of the NC DFR framework, the Abelian
gauge field associated with $U^{\star}(1)$ has acquired a mass
term. Besides, thanks to the NC scenario, some fields
are massive and others massless. Also in DFR scenario
we have obtained 3-line and 4-line vertex interactions and
the renormalizability of the model was preserved. The
residual symmetry $U^{\star}(1)$ was eliminated via the use of
the Higgs sector.
Moreover, we have introduced a second Higgs sector in
order to break the electroweak symmetry.   The masses
of both the old and new bosons were computed with the NC
contributions. Since the Weinberg angle was identified as
the basic angle to calculate the masses of the $W^{\pm}$ and
$Z^{0}$, we have used the experimental value of the sine of
the Weinberg angle in order to calculate the $W^{\pm}$ and $Z^{0}$
masses in a NC scenario. We have used the lower bound
for the first SSB scale given by $u = 1 \, \mbox{TeV}$. Finally, we
have obtained the masses for new antisymmetric bosons
of the DFR framework.

With the values of the charges well defined, and the relation between $\alpha$ and $\theta_{W}$ mixing angles,
we can list just below all the possible interactions between leptons, neutrinos and gauge bosons fields:
\\
\\
a) Interactions between leptons + neutrinos and the charged bosons $W^{\pm}$:
\begin{equation}\label{LnuW+-l}
{\cal L}_{\nu_{\ell} \, \ell-W^{\pm}}=-\frac{g_{1}}{2\sqrt{2}} \, \bar{\nu_{\ell}} \star \left(1+\gamma_{5}\right) \, \slash{\!\!\!W}^{+} \star\ell
-\frac{g_{1}}{2\sqrt{2}} \, \bar{\ell} \star \left(1+\gamma_{5}\right) \, \slash{\!\!\!W}^{-}  \star \nu_{\ell} \; .
\end{equation}
b) Interactions between leptons + neutrinos and photons:
\begin{eqnarray}\label{Lllgamma}
{\cal L}_{\bar{\ell}\ell-\gamma}&=&e \, \bar{\ell} \star \, \slash{\!\!\!\!A} \star \ell
+ \frac{e}{4} \, \, \bar{\ell} \star \gamma^{\mu} \left(1+\gamma^{5}\right)
\left[ \!\!\! \phantom{\frac{1}{2}} A_{\mu} \, , \, \ell \, \right]_{\star} ,
\end{eqnarray}
\begin{eqnarray}\label{LnunuGamma}
{\cal L}_{\bar{\nu}_{\ell}\nu_{\ell}-\gamma}=\frac{3e}{4} \, \bar{\nu}_{\ell}\star \gamma^{\mu} \left(1-\gamma_{5}\right) \left[ \!\!\! \phantom{\frac{1}{2}} \nu_{\ell} \, , \, A_{\mu} \, \right]_{\star} \; .
\end{eqnarray}
c) Interactions between the leptons + neutrinos and neutral boson-$Z$:
\begin{eqnarray}\label{LllZ}
{\cal L}_{\bar{\ell}\ell-Z}&=& \frac{g_{1}}{4\, \cos\theta_{W}} \, \bar{\ell} \star
\left(1-4\sin^{2}\theta_{W}+\gamma_{5} \right) \, \slash{\!\!\!\!Z} \star \ell
\nonumber \\
&& + \frac{g_{1}}{4\, \cos\theta_{W}} \, \bar{\ell}\star \gamma^{\mu} \left(1+\gamma_{5} \right) \left[ \!\!\! \phantom{\frac{1}{2}} \ell \, , \, Z_{\mu} \, \right]_{\star} \; ,
\end{eqnarray}
\begin{eqnarray}\label{LnunuZ}
{\cal L}_{\bar{\nu}_{\ell}\nu_{\ell}-Z}&=&\frac{g_{1}}{4 \, \cos\theta_{W}} \, \bar{\nu}_{\ell}\star \left(1+\gamma_{5}\right)\,
\slash{\!\!\!\!Z} \star \nu_{\ell}
\nonumber \\
&&
\hspace{-0.5cm} -\frac{3e}{4} \, \tan\theta_{W} \, \bar{\nu}_{\ell}\star  \gamma^{\mu}\left(1-\gamma_{5}\right)\left[ \!\!\! \phantom{\frac{1}{2}} \nu_{\ell} \, , \, Z_{\mu} \, \right]_{\star} \; .
\end{eqnarray}
d) Interactions between the leptons + neutrinos and neutral boson-$G$:
\begin{eqnarray}\label{LllG}
&&{\cal L}_{\bar{\ell}\ell-G}=- \frac{1}{2} \, g_{1}\sec\alpha \, \bar{\ell} \star  \left[1+\frac{3}{2} \, \tan^2\theta_{W}+\gamma_{5}\left(1-\tan^2\theta_{W} \right) \right] \, \slash{\!\!\!\! G}  \star \ell
\nonumber \\
&&-\frac{1}{2} \, g_{1} \, \tan\theta_{W} \tan\alpha \, \bar{\ell}\star \gamma^{\mu}\left[1+\frac{3}{2} \sec\theta_{W}
-\gamma_{5}\left(1-\frac{1}{2} \sec\theta_{W} \right)\right] \left[ \!\!\! \phantom{\frac{1}{2}} \ell \, , \, G_{\mu} \right]_{\star} ,
\hspace{1cm}
\end{eqnarray}
\begin{eqnarray}\label{LnunuG}
{\cal L}_{\bar{\nu}_{\ell}\nu_{\ell}-G}\!\!\!&&=-\frac{1}{2} \, g \sec\alpha \left(1+\frac{1}{2} \tan^2\theta_{W} \right)
\bar{\nu}_{\ell}\star \left( 1+\gamma_{5} \right) \, \slash{\!\!\!\! G}  \star \nu_{\ell}
\nonumber \\
&&-\frac{3}{4} \, g_{1} \, \tan\theta_{W} \, \tan\alpha \, \bar{\nu}_{\ell} \star \gamma^{\mu}\left(1-\gamma_{5}\right)\left[ \!\!\! \phantom{\frac{1}{2}} \nu_{\ell} \, , \, G_{\mu} \, \right]_{\star} .
\end{eqnarray}

%
%

The interactions list between the bosons $W^{\pm}$, $Z^{0}$, $A^{\mu}$ and $G^{\mu}$ is a tedious list of terms.   So we can represent here
the interactions $W^{\pm}$-$A^{\mu}$ and $W^{\pm}$-$Z$ to compare it with the commutative electroweak case:
%
%
\begin{figure}[!h]
\begin{center}
\newpsobject{showgrid}{psgrid}{subgriddiv=1,griddots=10,gridlabels=6pt}
\begin{pspicture}(-1,1.2)(14,3.3)
\psset{arrowsize=0.2 2}
\psset{unit=1}
%
%
%
\pscoil[coilarm=0,coilaspect=0,coilwidth=0.2,coilheight=1.0,linecolor=black](-1,1)(2,1)
\pscoil[coilarm=0,coilaspect=0,coilwidth=0.2,coilheight=1.0,linecolor=black](0.45,1.1)(0.45,3.1)
\put(0.7,2.8){\large$\gamma$}
\put(-0.95,1.2){\large$W^{+}$}
\put(1.7,1.2){\large$W^{-}$}
%
%
%
\pscoil[coilarm=0,coilaspect=0,coilwidth=0.2,coilheight=1.0,linecolor=black](3,1)(6,1)
\pscoil[coilarm=0,coilaspect=0,coilwidth=0.15,coilheight=1.0,linecolor=black](4.45,1.1)(3.5,3)
\pscoil[coilarm=0,coilaspect=0,coilwidth=0.15,coilheight=1.0,linecolor=black](4.45,1.1)(5.5,3)
\put(3.1,2.8){\large$\gamma$}
\put(5.7,2.8){\large$\gamma$}
\put(2.9,1.2){\large$W^{+}$}
\put(5.8,1.2){\large$W^{-}$}
%
%
\pscoil[coilarm=0,coilaspect=0,coilwidth=0.2,coilheight=1.0,linecolor=black](7,1)(10,1)
\pscoil[coilarm=0,coilaspect=0,coilwidth=0.2,coilheight=1.0,linecolor=black](8.45,1.1)(8.45,3.1)
\put(8.7,2.8){\large$Z^{0}$}
\put(6.9,1.2){\large$W^{+}$}
\put(9.7,1.2){\large$W^{-}$}
%
%
%
\pscoil[coilarm=0,coilaspect=0,coilwidth=0.2,coilheight=1.0,linecolor=black](11,1)(14,1)
\pscoil[coilarm=0,coilaspect=0,coilwidth=0.15,coilheight=1.0,linecolor=black](12.45,1.1)(13.5,3)
\pscoil[coilarm=0,coilaspect=0,coilwidth=0.15,coilheight=1.0,linecolor=black](12.45,1.1)(11.5,3)
\put(10.9,2.8){\large$Z^{0}$}
\put(13.6,2.8){\large$Z^{0}$}
\put(10.9,1.2){\large$W^{+}$}
\put(13.8,1.2){\large$W^{-}$}
\end{pspicture}
%
%
%
\end{center}
\end{figure}
%
%
\begin{eqnarray}
{\cal L}_{\gamma-W^{\pm}}=-\frac{i\,g_{1}}{2} \, \sin\theta_{W} \left(\partial_{\mu}W_{\nu}^{+}-\partial_{\nu}W_{\mu}^{+}\right)
\left\{W^{\mu-},A^{\nu} \right\}_{\star}
\nonumber \\
+\frac{i\,g_{1}}{2} \, \sin\theta_{W}  \left(\partial_{\mu}W_{\nu}^{-}-\partial_{\nu}W_{\mu}^{-}\right)
\left\{W^{\mu+},A^{\nu} \right\}_{\star}
\nonumber \\
-\frac{i\,g_{1}}{2} \, \sin\theta_{W} \left(\partial_{\mu}A_{\nu}-\partial_{\nu}A_{\mu}\right)
\left\{W^{\mu+},W^{\nu-} \right\}_{\star}
\nonumber \\
-i g_{1} \sin\theta_{W} \left(\partial_{\mu}W_{\nu}^{+}-\partial_{\nu}W_{\mu}^{+}\right)
\left[W^{\mu-},A^{\nu} \right]_{\star}
\nonumber \\
-i g_{1} \sin\theta_{W} \left(\partial_{\mu}W_{\nu}^{-}-\partial_{\nu}W_{\mu}^{-}\right)
\left[W^{\mu+},A^{\nu} \right]_{\star}
\nonumber \\
-\frac{i\,g_{1}}{4} \sin\theta_{W} \left(\partial_{\mu}A_{\nu}-\partial_{\nu}A_{\mu}\right)
\left[W^{\mu+},W^{\nu-} \right]_{\star} \; ,
\end{eqnarray}
\begin{eqnarray}
{\cal L}_{\gamma\gamma-W^{\pm}}\!&=&\!-\frac{g_{1}^{2}}{4} \, \sin^{2}\theta_{W} \left(
\frac{1}{2} \, \left\{W_{\mu}^{+},A_{\nu}\right\} -\frac{1}{2} \, \left\{ A_{\mu}, W_{\nu}^{\, +} \right\}_{\star}
-\left[W_{\mu}^{\, +},A_{\nu}\right]-\left[A_{\mu}, W_{\nu}^{\, +}\right] \phantom{\frac{1}{2}} \!\!\!\!\! \right) \, \times
\nonumber \\
&&\times \, \left(
\frac{1}{2} \, \left\{W^{\mu \, -},A^{\nu}\right\} -\frac{1}{2} \, \left\{ A^{\mu}, W^{\nu \, -} \right\}_{\star}
+\left[W^{\mu \, -},A^{\nu}\right]_{\star}+\left[A^{\mu}, W^{\nu \, -}\right]_{\star} \phantom{\frac{1}{2}} \!\!\!\!\! \right)
\nonumber \\
&& + \, g_{1}^{2} \, \sin^2\theta_{W} \, \left\{W_{\mu}^{+},W_{\nu}^{\,-}\right\}_{\star}
\left[A^{\mu},A^{\nu} \right]_{\star} \; ,
\end{eqnarray}
\begin{eqnarray}
{\cal L}_{Z^{0}-W^{\pm}}=-\frac{i\,g_{1}}{2} \, \cos\theta_{W} \left(\partial_{\mu}W_{\nu}^{+}-\partial_{\nu}W_{\mu}^{+}\right)
\left\{W^{\mu-},Z^{\nu} \right\}_{\star}
\nonumber \\
+\frac{i\,g_{1}}{2} \, \cos\theta_{W}  \left(\partial_{\mu}W_{\nu}^{-}-\partial_{\nu}W_{\mu}^{-}\right)
\left\{W^{\mu+},Z^{\nu} \right\}_{\star}
\nonumber \\
-\frac{i\,g_{1}}{2} \, \cos\theta_{W} \left(\partial_{\mu}Z_{\nu}-\partial_{\nu}Z_{\mu}\right)
\left\{W^{\mu+},W^{\nu-} \right\}_{\star}
\nonumber \\
+i g_{1} \, \sin^2\theta_{W} \sec\theta_{W} \left(\partial_{\mu}W_{\nu}^{+}-\partial_{\nu}W_{\mu}^{+}\right)
\left[W^{\mu-},Z^{\nu} \right]_{\star}
\nonumber \\
+i g_{1} \, \sin^2\theta_{W} \sec\theta_{W} \left(\partial_{\mu}W_{\nu}^{-}-\partial_{\nu}W_{\mu}^{-}\right)
\left[W^{\mu+},Z^{\nu} \right]_{\star}
\nonumber \\
+\frac{i\,g_{1}}{4} \, \sin^2\theta_{W} \sec\theta_{W} \left(\partial_{\mu}Z_{\nu}-\partial_{\nu}Z_{\mu}\right)
\left[W^{\mu+},W^{\nu-} \right]_{\star} \; ,
\end{eqnarray}
\begin{eqnarray}
{\cal L}_{Z^{0}Z^{0}-W^{\pm}}=-\frac{g_{1}^{2}}{4} \, \cos^{2}\theta_{W} \left(
\frac{1}{2} \, \left\{W_{\mu}^{+},Z_{\nu}\right\} -\frac{1}{2} \, \left\{ Z_{\mu}, W_{\nu}^{\, +} \right\}_{\star}
\right.
\nonumber \\
\left.
- \tan^2\theta_{W}\left[W_{\mu}^{\, +},Z_{\nu}\right]_{\star} - \tan^2\theta_{W}\left[Z_{\mu}, W_{\nu}^{\, +}\right]_{\star} \phantom{\frac{1}{2}} \!\!\!\!\! \right) \, \times
\nonumber \\
\times \, \left(
\frac{1}{2} \, \left\{W^{\mu \, -},Z^{\nu}\right\} -\frac{1}{2} \, \left\{ Z^{\mu}, W^{\nu \, -} \right\}_{\star}
\right.
\nonumber \\
\left.
+ \tan^2\theta_{W}\left[W^{\mu \, -},Z^{\nu}\right]_{\star}+ \tan^2\theta_{W}\left[Z^{\mu}, W^{\nu \, -}\right]_{\star} \phantom{\frac{1}{2}} \!\!\!\!\! \right)
\nonumber \\
- g_{1}^{2} \, \sin^2\theta_{W} \left\{W_{\mu}^{+},W_{\nu}^{\,-}\right\}_{\star}
\left[Z^{\mu},Z^{\nu} \right]_{\star} \; .
\end{eqnarray}

In the commutative limit, all the commutators $[,]_{\star}$ go to zero, and the previous interactions can be reduced
to the rules of the usual GSW electroweak model.
\\
\\
\hspace{0.5cm}
The interactions of $3$-line vertex emerging from (\ref{LYukawa1}) are given by:
\\
\\
a) Interaction between the leptons, neutrinos and charged bosons $B^{\pm}$
\begin{equation}\label{LlnuB}
{\cal L}_{Yukawa-\ell\nu_{\ell}- B^{\pm}}^{\, (1)}=-\frac{g_{2}}{4\sqrt{2}} \, \bar{\ell} \left(1-\gamma_{5}\right) \star (\overrightarrow{\Sigma}
\cdot \overrightarrow{B}^{\, -} ) \star \nu_{\ell}
-\frac{g_{2}}{4\sqrt{2}} \, \bar{\nu}_{\ell} \left(1-\gamma_{5}\right) \star (\overrightarrow{\Sigma} \cdot \overrightarrow{B}^{\, +} ) \star \ell \; .
\end{equation}
b) Interaction between leptons and the Higgs-One:
\begin{eqnarray}\label{LlH1}
{\cal L}_{Yukawa-\bar{\ell}\ell- H_{1}}^{\, (1)}=i \, u^{-1} \, \bar{\ell} \star H_{1} \star \lambda \, \overrightarrow{\Sigma}\cdot \overrightarrow{\nabla}_{\theta} \, \ell
-i \, u^{-1} \, \lambda \,  \overrightarrow{\nabla}_{\theta}\bar{\ell} \star H_{1} \star \cdot\overrightarrow{\Sigma} \, \ell \; .
\end{eqnarray}
c) Interaction between leptons and neutral bosons $A\,'_{i}$:
\begin{eqnarray}\label{LllAmunu}
{\cal L}_{Yukawa-\bar{\ell}\ell- A\,'_{i}}^{\, (1)}\!\!&=\!\!&
\frac{3}{4} \, g_{2}\sin\theta_{2} \, \bar{\ell}\star \overrightarrow{\Sigma}\cdot \overrightarrow{A}' \star\ell
+\frac{1}{2} \, g^{\prime} \cos\theta_{2} \, \bar{\ell}\star \overrightarrow{\Sigma}\cdot \overrightarrow{A}' \star\ell \,
\nonumber \\
&&
-\frac{1}{4} \, g^{\prime} \, \cot\beta \, \left(\frac{3}{2} + \cot \beta \,  \right) \, \bar{\ell} \left(1+\gamma^{5}\right) \star \overrightarrow{\Sigma} \cdot \left[ \, \ell \, , \, \overrightarrow{A}' \, \right]_{\star}
\nonumber \\
&&
-\frac{1}{4} \, g^{\prime} \, \cot\beta \, \left(\frac{3}{2} + \cot \beta \,  \right) \, \left[ \, \bar{\ell} \, , \, A\,'_{i} \, \right]_{\star} \left(1-\gamma^{5}\right) \star \Sigma_{i} \, \ell \; .
\end{eqnarray}
d) Interaction between leptons and the neutral bosons $Z\,'_{i}$:
\begin{eqnarray}\label{LllZmunu}
{\cal L}_{Yukawa-\bar{\ell}\ell- Z\,'_{i}}^{\, (1)}&=&
\frac{1}{4} \, g_{2} \cos\theta_{2} \left( \, 1 - 6 \, \sin^2 \beta \, \right) \bar{\ell} \star \overrightarrow{\Sigma}\cdot \overrightarrow{Z} \star\ell
\nonumber \\
&&
-\frac{1}{4} \, g^{\prime} \, \sin\theta_{2} \, \bar{\ell} \left(1+\gamma^{5}\right) \star \Sigma^{i}\left[ \, \ell \, , \, Z_{i} \, \right]_{\star}
\nonumber \\
&&
-\frac{1}{4} \, g^{\prime} \, \left(\frac{1}{2} + \cot \beta  \right) \sin\theta_{2} \, \bar{\ell} \left(1-\gamma^{5}\right) \star \Sigma_{i}
\left[ \, \ell \, , \, Z_{i} \, \right]_{\star}
\nonumber \\
&&
-\frac{1}{4} \, g^{\prime}\sin\theta_{2} \, \left[ \, \bar{\ell} \, , \, Z_{i} \, \right]_{\star} \left(1-\gamma^{5}\right) \star \Sigma_{i} \, \ell
\nonumber \\
&&
-\frac{1}{4} \, g^{\prime} \, \left(\frac{1}{2} + \cot \beta  \right)  \sin\theta_{2} \, \left[ \, \bar{\ell} \, , \, Z_{i} \, \right]_{\star} \left(1+\gamma^{5}\right) \star \Sigma_{i} \, \ell
\; .
\end{eqnarray}
e) Interaction between leptons and the neutral bosons $G\,'_{i}$:
\begin{eqnarray}\label{LllGmunu}
{\cal L}_{Yukawa-\bar{\ell}\ell- G\,'_{i}}^{\, (1)}&=&
\frac{5}{8} \, g_{2} \, \sin\theta_{2} \, \bar{\ell}\star \overrightarrow{\Sigma}\cdot \overrightarrow{G}'  \star\ell
\nonumber \\
&&
+\frac{1}{4} \, g^{\prime} \, \cos\theta_{2} \, \bar{\ell} \left(1+\gamma^{5}\right) \star \Sigma_{i} \left[ \, \ell \, , \, G\,'_{i} \, \right]_{\star}
\nonumber \\
&&
+\frac{1}{4} \, g^{\prime} \, \left(\frac{1}{2} + \cot \beta  \right) \cos\theta_{2} \, \bar{\ell} \left(1-\gamma^{5}\right) \star \Sigma_{i} \, \left[ \, \ell \, , \, G\,'_{i} \,  \right]_{\star}
\nonumber \\
&&
+\frac{1}{4} \, g^{\prime}\cos\theta_{2} \, \left[ \, \bar{\ell} \, , \, G\,'_{i} \, \right]_{\star} \left(1-\gamma^{5}\right) \star \Sigma_{i} \, \ell
\nonumber \\
&&
+\frac{1}{4} \, g^{\prime} \, \left(\frac{1}{2} + \cot \beta  \right) \cos\theta_{2} \, \left[ \, \bar{\ell} \, , \, G\,'_{i} \, \right]_{\star} \left(1+\gamma^{5}\right) \star \Sigma_{i} \, \ell
\; .
\end{eqnarray}

The Yukawa Lagrangian also produces interactions of $4$-line vertex, where we will write some terms just below
\begin{eqnarray}\label{LYukawa4}
{\cal L}_{Yukawa-4}^{\, (1)}&=&
-\frac{g_{2} \, u^{-1}}{4\sqrt{2}} \, \bar{\ell} \star \left(1-\gamma_{5}\right) H_{1} \star \overrightarrow{\Sigma}\cdot \overrightarrow{B}^{-} \star \nu_{\ell}
\nonumber \\
&&
-\frac{g_{2} \, u^{-1}}{4\sqrt{2}} \, \bar{\nu}_{\ell} \star \left(1-\gamma_{5}\right) \overrightarrow{\Sigma}\cdot \overrightarrow{B}^{+} \star H_{1} \star \ell
\nonumber \\
&&
-\frac{g^{\prime} \, u^{-1}}{2 \, \tan\beta} \, \, \bar{\ell}\star\Sigma_{i} \, \left\{ \, A\,'_{i} \, , \, H_{1} \, \right\}_{\star}\star\ell
\nonumber \\
&&
-\frac{g^{\prime} \, u^{-1}}{2 \, \tan\beta} \, \, \bar{\ell}\star \gamma^{5} \, \Sigma_{i} \, \left[ \, A\,'_{i} \, , \, H_{1} \, \right]_{\star}\star\ell
+ \ldots
\end{eqnarray}
%

The $4$-line vertex have the coupling constants with length dimension, which depends on $u^{-1}$. We can compute approximately
the scale of this VEV as being $u \sim 1-10 \; \mbox{TeV}$, so we can affirm that the coupling constants emerging from (\ref{LYukawa4}) are extremely weak. Therefore, the contribution of those interactions must be weaker in comparison to others ones that, in the commutative limit go to zero. The renormalizability of the model is a puzzle that is out of the scope of this paper, and it deserves a
detailed attention.   It is an ongoing research.

\section{The gauge symmetry after SSBs}
\renewcommand{\theequation}{7.\arabic{equation}}
\setcounter{equation}{0}

Now we will see which symmetry remains in the model after the electroweak spontaneous breaking.
To obtain the gauge transformations of the new fields $G^{\mu}-Y^{\mu}$, we will write the elements of the NC Abelian groups as
\begin{eqnarray}
V_{1}(x,\theta)=e^{i \, g_{1} \, \eta(x,\theta)}
\hspace{0.5cm} \mbox{and} \hspace{0.5cm}
V_{2}(x,\theta)=e^{i \, J \, g_{1}^{\prime} \, \xi(x,\theta)} \; ,
\end{eqnarray}
where $\eta$ and $\xi$ are real functions. The infinitesimal transformations of the NC vector Abelian fields are given by
\begin{eqnarray}
A_{\mu}^{0} \longmapsto A_{\mu}^{0 \, \prime} &=& A_{\mu}^{ \, 0} + i \, \left[ \, g_{1} \, \eta \, , \, A_{\mu}^{\, 0} \, \right]_{\star}
-\partial_{\mu}\eta \; ,
\nonumber \\
B_{\mu} \longmapsto B_{\mu}^{\, \prime} &=& B_{\mu} + i \, \left[ \, J \, g_{1}^{\prime} \, \xi \, , \, B_{\mu} \, \right]_{\star}
-\partial_{\mu}\xi \; .
\end{eqnarray}
Since $Y=0$ for neutral bosons, we can obtain here $J=1$ from the relation (\ref{YJ}).  Establishing that $g_{1}\, \eta = g_{1}^{\prime} \, \xi$, redefining $\eta(x,\theta)=\sin\alpha \, f(x,\theta)$ and $\xi(x,\theta)=\cos\alpha \, f(x,\theta)$, the relation (\ref{gg1g1prime}) gives us the $U_{Y}(1)$
transformation
\begin{eqnarray}\label{transfYAmunu}
G_{\mu} \longmapsto G_{\mu}^{\prime} &=& G_{\mu} + \frac{i}{2} \, \left[ \, g \, f(x,\theta) \, , \, G_{\mu} \, \right]_{\star} \; ,
\\
Y_{\mu} \longmapsto Y_{\mu}^{\, \prime} &=& Y_{\mu} + \frac{i}{2} \, \left[ \, g \, f(x,\theta) \, , \, Y_{\mu} \, \right]_{\star}
- \partial_{\mu}f(x,\theta) \; .
\end{eqnarray}
These are typical transformations of a NC Abelian group, where
$G_{\mu}$ is a NC massive field, and $Y_{\mu}$ is the NC massless field of $U_{Y}^{\star}(1)$. After the first
SSB, we recall that the infinitesimal gauge transformation of the vector bosons $A_{\mu}^{ \, a}$ , under
$SU_{L}^{\star}(2)_{A^{\mu a}}$, is given by
\begin{eqnarray}\label{transfAmuinf}
A_{\mu}^{\prime \, a} &=& A_{\mu}^{\,a}-\frac{1}{2} \,\varepsilon^{abc}\left\{ \, g_1 \, \omega^{b} \, , \, A_{\mu}^{c} \, \right\}_{\star}
+\frac{i}{2} \, \left[ \, g \, f \, , \, A_{\mu}^{a} \, \right]_{\star}-\partial_{\mu}\omega^{a} \; ,
\end{eqnarray}
where the mixing between the $3$-component of (\ref{transfAmuinf}) and the gauge transformation of $Y_{\mu}$ will give us the result of both the transformations of $Z^{0}$ and the massless photon $A^{\mu}$. We fix the parameters $\omega^{1}=\omega^{2}=0$,
and we will use Eqs. (\ref{transfAYZ}), and the transformations of $Z^{0}$ and $A^{\mu}$-photon can be written as
\begin{eqnarray}\label{transfinfgaugeZAomega3f}
Z_{\mu} \longmapsto Z_{\mu}^{\prime} &=& Z_{\mu}
+\frac{i}{2} \, \left[ \, g \, f(x,\theta) \, , \, Z_{\mu} \, \right]_{\star}-\partial_{\mu}\left(\omega^{3}\cos\theta_{W}-f \, \sin\theta_{W} \right) \; ,
\nonumber \\
A_{\mu} \longmapsto A_{\mu}^{\prime}&=& A_{\mu}
+\frac{i}{2} \, \left[ \, g \, f(x,\theta) \, , \, A_{\mu}\right]_{\star}-\partial_{\mu}\left(\omega^{3}\cos\theta_{W}+f \, \sin\theta_{W}  \right) \; .
\end{eqnarray}
Hence, let us write $\omega^{3}\cos\theta_{W}=f \, \sin\theta_{W}$, and redefining $f(x,\theta)=\cos\theta_{W} \, \Lambda(x,\theta)$,
we have that
\begin{eqnarray}\label{transfinfgaugeZA}
Z_{\mu} \longmapsto Z_{\mu}^{\prime} &=& Z_{\mu}
+\frac{i}{2} \, \left[ \, e \, \Lambda(x,\theta) \, , \, Z_{\mu} \, \right]_{\star} \; ,
\nonumber \\
A_{\mu} \longmapsto A_{\mu}^{\prime}&=& A_{\mu}
+\frac{i}{2} \, \left[ \, e \, \Lambda(x,\theta) \, , \, A_{\mu} \, \right]_{\star}-\partial_{\mu}\Lambda(x,\theta) \; ,
\end{eqnarray}
and for the charged bosons $W^{\pm}$, we can obtain that
\begin{eqnarray}
W_{\mu}^{+} \longmapsto W_{+}^{\prime \, +} &=& W_{\mu}^{+}  + i \, e \, \Lambda(x,\theta) \star  W_{\mu}^{+} \; ,
\nonumber \\
W_{\mu}^{-} \longmapsto W_{\mu}^{\prime \, -} &=& W_{\mu}^{-}  - i \, e \, W_{\mu}^{-} \star \Lambda(x,\theta)
\; .
\end{eqnarray}
Therefore, the exact gauge transformations for the vector bosons are
\begin{eqnarray}\label{transfG}
G_{\mu} \longmapsto G_{\mu}^{\prime}&=&U_{\Lambda} \star G_{\mu} \star U_{\Lambda}^{-1} \; ,
\nonumber \\
Z_{\mu} \longmapsto Z_{\mu}^{\prime}&=&U_{\Lambda} \star Z_{\mu} \star U_{\Lambda}^{-1} \; ,
\nonumber \\
A_{\mu} \longmapsto A_{\mu}^{\, \prime}&=& U_{\Lambda} \star A_{\mu} \star U_{\Lambda}^{-1}-\frac{i}{e/2} \, U_{\Lambda} \star \partial_{\mu}U_{\Lambda}^{-1} \; ,
\nonumber \\
W_{\mu}^{+} \longmapsto W_{\mu}^{\prime \, +} &=& e^{\, i \, e \, \Lambda(x,\theta)} \star W_{\mu}^{+}  \; ,
\nonumber \\
W_{\mu}^{-} \longmapsto W_{\mu}^{\prime \, -} &=& W_{\mu}^{-} \star e^{-i \, e \, \Lambda(x,\theta)} \; .
\end{eqnarray}
where $U_{\Lambda}(x,\theta)=e^{\frac{i}{2} \, e \, \Lambda(x,\theta)}$ is the element of the residual NC electromagnetic group $U_{em}^{\star}(1)$.
Analogously, the NC tensor fields sector have the transformations
\begin{eqnarray}\label{transfBBX}
B_{i}^{\, 0} \longmapsto B_{i}^{ \, 0 \, \prime}&=& B_{i}^{\, 0} + i \, \left[ \, g_{1} \, \eta \, , \, B_{i}^{\, 0} \, \right]_{\star}
-\lambda \, \partial_{\theta \, i}\eta \; ,
\nonumber \\
B_{i}^{ \, a} \longmapsto B_{i}^{ \, a \, \prime}&=& B_{i}^{\, a}
-\frac{1}{2} \, \varepsilon^{abc} \left\{ \, g_{1} \, \omega^{b} \, , \, B_{i}^{\, c} \, \right\}_{\star}
+ i \, \left[ \, g_{1} \, \eta \, , \, B_{i}^{\, a} \, \right]_{\star}
-\lambda \partial_{\theta \, i}\omega^{a} \; ,
\nonumber \\
C_{i} \longmapsto C_{i}^{\, \prime} &=& C_{i} + i \, \left[ \, J \, g_{1}^{\prime} \, \xi \, , \, C_{i} \, \right]_{\star}
- \frac{g_{1}^{\prime}}{g_{2}^{\prime}} \, \lambda \partial_{\theta \, i}\xi \; ,
\end{eqnarray}
using the mixing (\ref{transfB0X}) and the previous conditions, we obtain
\begin{eqnarray}\label{transfYAmunu}
G'_{i} \longmapsto \left(G'_{i}\right)^{\prime} &=& G'_{i} + \frac{i}{2} \, \left[ \, g \, f(x,\theta) \, , \, G'_{i} \, \right]_{\star}
\; ,
\\
Y'_{i} \longmapsto \left(Y'_{i}\right)^{\, \prime} &=& Y'_{i} + \frac{i}{2} \, \left[ \, g \, f(x,\theta) \, , \, Y'_{i} \, \right]_{\star}
- \frac{g}{g^{\prime}} \, \lambda \, \partial_{\theta \, i}f(x,\theta) \; .
\end{eqnarray}
Using the previous conditions on the parameters, we can obtain the transformations of the tensor fields
%
%
%
\begin{eqnarray}
G'_{i} \longmapsto \left(G'_{i}\right)^{\, \prime} &=& U_{\Lambda} \star G'_{i} \star U_{\Lambda}^{-1} \; ,
\nonumber \\
Z'_{i} \longmapsto \left(Z'_{i}\right)^{\prime} &=& U_{\Lambda} \star Z'_{i} \star U_{\Lambda}^{-1} \; ,
\nonumber \\
A'_{i} \longmapsto \left(A'_{i}\right)^{\, \prime} &=& A'_{i} + \frac{i}{2} \, \left[ \, e \, \Lambda(x,\theta) \, , \, A'_{i} \, \right]_{\star}
- \frac{\sin\theta_{W}}{2 \, \sin\theta_{2}} \, \lambda \, \partial_{\theta \, i} \Lambda(x,\theta) \; ,
\nonumber \\
B_{i}^{\, +} \longmapsto B_{i}^{\prime \, +} &=& e^{\, i \, e \, \Lambda(x,\theta)} \star B_{i}^{\, +}  \; ,
\nonumber \\
B_{i}^{\, -} \longmapsto B_{i}^{\prime \, -} &=& B_{i}^{\, -} \star e^{-i \, e \, \Lambda(x,\theta)} \; .
\end{eqnarray}
After the gauge SSB, we have obtained four massive charged bosons $W^{\pm}$ and $B^{\pm}$,
plus four massive neutral bosons $Z^{0}$, $G^{\mu}$, $Z'_{i}$, $G'_{i}$
and two massless neutral bosons $A^{\mu}$ and $A'_{i}$. It explains the final gauge symmetry
$U_{em}^{\star}(1) \times U^{\star}(1)_{A'_{i}}$.
%

%
%
%


%
\section{Conclusions and perspectives}
The analysis of theories or models that unify quantum mechanics and general relativity are one of the great challenges in theoretical physics.   In this way, the idea that the introduction of an uncertainty in the position coordinate thanks to the gravitational field introduces a NC spacetime framework that has been a target of intense investigation through the last few years. Following a different basic idea and another motivation, Snyder was the pioneer with the first published paper about this issue in order to eliminate the infinities that dwell QFT. However, his hopes were broken and a NC spacetime subject was put to sleep for more than forty years. The results obtained in string theory have ignited a new era in the NC literature.

However, the spacetimes where the NC parameter is a constant one have Lorentz invariance problems.  One of the solutions is to promote the NC parameter to a position coordinate.   This was the idea of Doplicher, Fredenhagen and Roberts.   An an extended Hilbert space with ten dimensions (four Minkowskian and six $\theta$ coordinates) was constructed. Recently, it was demonstrated \cite{jhep} precisely that the formalism has a conjugate momentum for the $\theta$ coordinate and several elements of a DFR QFT were constructed \cite{EMCAbreuMJNeves2012,EMCAbreuMJNeves2013,EMCAbreuMJNeves2014}.   However, to avoid unitarity we will keep the time components of $\theta^{\mu\nu}$ equal to zero, namely, $\theta^{0i}=0$.  With these definitions, we have constructed a metric tensor for the DFR space-time.

In this way, we have introduced in this work, together with new numerical results, one more step in the formulation of a DFR QFT.  At this time, we have analyzed some elements of the NC standard model such as the electroweak standard model.
Since the position and $\theta$ coordinates are independent variables, the Weyl-Moyal product keeps its associative property and it is the underlying product, as usual in canonical NC models.

With this understanding in mind, we have introduced new ideas and concepts in DFR formalism and we began with the construction of the symmetry group $U^{\star}_{L} (2)_{A^{\mu}} \times U^{\star}_R (1)_{X^{\mu}} \times U^{\star}_{L} (2)_{B_{i}} \times U^{\star}_R (1)_{C_{i}} $, which is the DFR version of the Glashow-Salam-Weinberg model concerning the electroweak interaction, in order to introduce left and right-handed fermionic sectors. Some elements such as covariant derivatives, gauge transformations and gauge invariant Lagrangians were constructed and the interactions between leptons, neutrinos and gauge fields were discussed.

After that we have introduced the first Higgs sector to break one of the two Abelian NC symmetries in order to destroy the residual model's $U^*(1)$ symmetry. The spontaneous symmetry breaking was discussed and, in this way, the Higgs Lagrangian was introduced.   We have seen that in the context of the NC DFR framework, the Abelian gauge field associated with $U^*(1)$ have acquired a mass term.   Besides, thanks to the NC scenario, some fields are massive and others, massless.   Also in the NC context we have obtained 3-line and 4-line vertex interactions and the renormalizability of the model was preserved.   We have seen that the coupling constants emerging from the 4-line Yukawa Lagrangian are extremely weak, i.e., weaker in comparison to other ones.   The residual symmetry $U^* (1)$ was eliminated via the use of the Higgs sector.

Moreover, we have introduced a second Higgs sector in order to break the electroweak symmetry and the masses of the new bosons were computed with the NC contributions.  Since the Weinberg angle was identified as the basic angle to calculate the masses of the $W^{\pm}$ and $Z^0$, we have used the experimental value of the sine of the Weinberg angle in order to calculate the $W^{\pm}$ and $Z^0$ masses in an NC scenario.  We have used the lower bound for the NC parameter given by $1$ GeV. The mass of $G'_{i}$ can be estimated in terms of the mixing angle-$\beta$, but it was not calculated here.
This mass must be computed using a vertex interaction of leptons with $G'_{i}$ - propagator at one loop approximation. It is a motivation for an ongoing research that will be published elsewhere. Finally, we have discussed the DFR gauge transformations symmetries.

Since we have discussed the main elements of a standard electroweak model in a NC DFR context, the next step would be the introduction of the other standard model elements in DFR formalism.

\section{Acknowledgments}

\ni E.M.C.A. thanks CNPq (Conselho Nacional de Desenvolvimento Cient\' ifico e Tecnol\'ogico), Brazilian scientific support federal agency, for partial financial support, Grants No. 301030/2012-0 and No. 442369/2014-0 and the hospitality of Theoretical Physics Department at Federal University of Rio de Janeiro (UFRJ), where part of this work was carried out. The authors thank also professor J. A. Helay\"el Neto, for valuable discussions.


\end{document}